\documentclass[twocolumn, trackchanges, twocolappendix]{aastex631}

\usepackage{ulem}
\usepackage{graphicx, amsmath, float, booktabs}
\usepackage{array}
\usepackage{placeins}
\usepackage{enumitem} 
\usepackage{threeparttable}
\usepackage{multirow}
\usepackage{tabu}
\usepackage{tabularx}
\usepackage{fancyhdr}
\usepackage{xcolor}
\usepackage{CJKutf8}

\newcommand{\TLchinesename}{{\begin{CJK}{UTF8}{gbsn}(李坦达)\end{CJK}}}
\newcommand{\QXchinesename}{{\begin{CJK}{UTF8}{gbsn}(熊强)\end{CJK}}}
\newcommand{\JYchinesename}{{\begin{CJK}{UTF8}{gbsn}(余杰)\end{CJK}}}
\newcommand{\ZLchinesename}{{\begin{CJK}{UTF8}{gbsn}(李志文)\end{CJK}}}
\newcommand{\SLchinesename}{{\begin{CJK}{UTF8}{gbsn}(毕少兰)\end{CJK}}}
\newcommand{\XZchinesename}{{\begin{CJK}{UTF8}{gbsn}(张先飞)\end{CJK}}}
\newcommand{\HBchinesename}{{\begin{CJK}{UTF8}{gbsn}(苑海波)\end{CJK}}}

\makeatletter
\newcommand*\fsize{\f@size pt\relax}
\makeatother
\usepackage{comment}

\newcommand{\kepler}[0]{\emph{Kepler}}

\newcommand{\tess}[0]{\emph{TESS}}

\newcommand{\gaia}[0]{\emph{Gaia}}

\newcommand{\Dnu}[0]{$\Delta\nu$}

\newcommand{\numax}[0]{$\nu_{\rm max}$}

\newcommand{\p}[0]{$P$}
\newcommand{\mk}[0]{$M_{K}$}
\newcommand{\mnumax}[0]{{\tt\string numax\_model}}
\newcommand{\mdnu}[0]{{\tt\string dnu\_model}}

\begin{document}

\thispagestyle{empty}

\title{Asteroseismology of Long-Period Variables with OGLE-IV data: Using
Global Seismic Parameters as Luminosity Indicators}

\correspondingauthor{Tanda Li}
\email{litanda@bnu.edu.cn}

\correspondingauthor{Jie Yu}
\email{Jie.Yu@anu.edu.au}

\author[0009-0009-5492-1343]{Qiang Xiong\QXchinesename}
\affiliation{School of Physics and Astronomy, Beijing Normal University, Beijing, 100875, China}

\author[0000-0001-6396-2563]{Tanda Li\TLchinesename}
\affiliation{School of Physics and Astronomy, Beijing Normal University, Beijing, 100875, China}
\affiliation{Institute for Frontiers in Astronomy and Astrophysics, Beijing Normal University, Beijing 102206, China}

\author[0000-0002-0007-6211]{Jie Yu\JYchinesename}
\affiliation{School of Computing, Australian National University, Acton, ACT 2601, Australia}

\author[0009-0003-3321-6393]{Zhiwen Li\ZLchinesename}
\affiliation{School of Physics and Astronomy, Beijing Normal University, Beijing, 100875, China}
\affiliation{Institute for Frontiers in Astronomy and Astrophysics, Beijing Normal University, Beijing 102206, China}

\author[0000-0002-7642-7583]{Shaolan Bi\SLchinesename}
\affiliation{School of Physics and Astronomy, Beijing Normal University, Beijing, 100875, China}
\affiliation{Institute for Frontiers in Astronomy and Astrophysics, Beijing Normal University, Beijing 102206, China}

\author[0000-0002-3672-2166]{Xianfei Zhang\XZchinesename}
\affiliation{School of Physics and Astronomy, Beijing Normal University, Beijing, 100875, China}
\affiliation{Institute for Frontiers in Astronomy and Astrophysics, Beijing Normal University, Beijing 102206, China}

\author[0000-0003-2471-2363]{Haibo Yuan\HBchinesename}
\affiliation{School of Physics and Astronomy, Beijing Normal University, Beijing, 100875, China}
\affiliation{Institute for Frontiers in Astronomy and Astrophysics, Beijing Normal University, Beijing 102206, China}

\author[0000-0002-7777-0842]{I. Soszy\'nski}
\affiliation{Astronomical Observatory, University of Warsaw, Al. Ujazdowskie 4, 00-478 Warszawa, Poland}

\author[0000-0001-5207-5619]{A. Udalski}
\affiliation{Astronomical Observatory, University of Warsaw, Al. Ujazdowskie 4, 00-478 Warszawa, Poland}

\author[0000-0002-0548-8995]{M. K. Szyma\'nski}
\affiliation{Astronomical Observatory, University of Warsaw, Al. Ujazdowskie 4, 00-478 Warszawa, Poland}

\author[0000-0001-9439-604X]{D. M. Skowron}
\affiliation{Astronomical Observatory, University of Warsaw, Al. Ujazdowskie 4, 00-478 Warszawa, Poland}

\author[0000-0002-2335-1730]{J. Skowron}
\affiliation{Astronomical Observatory, University of Warsaw, Al. Ujazdowskie 4, 00-478 Warszawa, Poland}

\author[0000-0002-2339-5899]{P. Pietrukowicz}
\affiliation{Astronomical Observatory, University of Warsaw, Al. Ujazdowskie 4, 00-478 Warszawa, Poland}

\author[0000-0002-9245-6368]{R. Poleski}
\affiliation{Astronomical Observatory, University of Warsaw, Al. Ujazdowskie 4, 00-478 Warszawa, Poland}

\author[0000-0003-4084-880X]{S. Koz\l{}owski}
\affiliation{Astronomical Observatory, University of Warsaw, Al. Ujazdowskie 4, 00-478 Warszawa, Poland}

\author[0000-0001-7016-1692]{P. Mr\'oz}
\affiliation{Astronomical Observatory, University of Warsaw, Al. Ujazdowskie 4, 00-478 Warszawa, Poland}

\author[0000-0001-6364-408X]{K. Ulaczyk}
\affiliation{Department of Physics, University of Warwick, Gibbet Hill Road, Coventry, CV4 7AL, UK}
\affiliation{Astronomical Observatory, University of Warsaw, Al. Ujazdowskie 4, 00-478 Warszawa, Poland}

\author[0000-0002-9326-9329]{K. Rybicki}
\affiliation{Department of Particle Physics and Astrophysics, Weizmann Institute of Science, Rehovot 76100, Israel}
\affiliation{Astronomical Observatory, University of Warsaw, Al. Ujazdowskie 4, 00-478 Warszawa, Poland}

\author[0000-0002-6212-7221]{P. Iwanek}
\affiliation{Astronomical Observatory, University of Warsaw, Al. Ujazdowskie 4, 00-478 Warszawa, Poland}

\author[0000-0002-3051-274X]{M. Wrona}
\affiliation{Astronomical Observatory, University of Warsaw, Al. Ujazdowskie 4, 00-478 Warszawa, Poland}

\author[0000-0002-1650-1518]{M. Gromadzki}
\affiliation{Astronomical Observatory, University of Warsaw, Al. Ujazdowskie 4, 00-478 Warszawa, Poland}

\author[0000-0002-8911-6581]{M. Mr\'oz}
\affiliation{Astronomical Observatory, University of Warsaw, Al. Ujazdowskie 4, 00-478 Warszawa, Poland}

\begin{abstract}

Long-period variables (LPVs) are high-luminosity red giants or supergiants with pulsation periods ranging from days to years. Many LPVs in the Large Magellanic Cloud (LMC) and Galactic Bulge (BLG) have been continuously observed over a time span of 26 years by the Optical Gravitational Lensing Experiment (OGLE) survey. Using OGLE-IV data, we applied Gaussian Processes with kernels tailored for solar-like oscillations to extract two global asteroseismic parameters: the frequency of maximum power (\numax) and the large frequency separation (\Dnu), for LPVs with primary mode periods (\p1) between 10 and 100 days in the LMC and BLG. We found that the \numax-\Dnu\ relation for LPVs in this work aligns with that of \mbox{lower-luminosity} \kepler\ red giants, confirming that the pulsations of these LPVs are likely solar-like. We found that \numax\ and \Dnu\ can serve as luminosity indicators. Compared to \p1, \numax\ and \Dnu\ exhibit significantly tighter correlations with the absolute magnitude in the 2MASS $K_s$ band (\mk), with corresponding scatter of 0.27 mag and 0.21 mag, respectively. Using the calibrated \numax-\mk\ and \Dnu-\mk\ relations for LPVs in the LMC, we determined the \mk\ values for individual stars in the BLG. By accounting for extinction, we further calculated the distances to 4,948 BLG stars. The peak of the resulting distance distribution corresponds to an estimated distance to the Galactic center of approximately 9.1 kpc, which appears to be overestimated, suggesting that the seismic luminosity relation calibrated from the LMC may not be directly applicable to BLG stars.\\

\noindent $Unified\ Astronomy\ Thesaurus\ concepts:$ \href{http://astrothesaurus.org/uat/73}{Asteroseismology (73)}; 
\href{http://astrothesaurus.org/uat/1930}{Gaussian Processes regression (1930)}; 
\href{http://astrothesaurus.org/uat/935}{Long-period variables (935)}; 
\href{http://astrothesaurus.org/uat/2123}{OGLE Small Amplitude Red Giants (2123)}; 
\href{http://astrothesaurus.org/uat/1625}{Stellar pulsations (1625)};

\end{abstract}

\section{Introduction} \label{sec:intro} 

Long-period variables (LPVs) are intrinsically luminous red giants or red supergiants, many of which are characterized by high-amplitude pulsations \citep[e.g.,][]{soszynski2009optical}, significant mass loss \citep[e.g.,][]{hofner2018}, and serve as crucial tools in astrophysics for tracing stellar populations and estimating astronomical distances \citep[e.g.,][]{2022A&A...658L...1T}. The study of LPV pulsations has continued for centuries, through the analysis of light curves from ground-based surveys, such as the Optical Gravitational Lensing Experiment \citep[OGLE,][]{1992AcA....42..253U}, the Massive Compact Halo Object project \citep[MACHO,][]{alcock1999calibration}, and the Asteroid Terrestrial-impact Last Alert System (ATLAS, \citealt{auge2020, hey2024}), as well as from space-based telescopes like \kepler\ \citep[e.g., ][]{mosser_period-luminosity_2013, stello2014a, yu2020asteroseismology} and \gaia\ \citep{lebzelter2023}.

LPVs exhibit pulsations with periods ranging from days to years. Among them, Mira variables are typically characterized by high-amplitude, single-mode pulsations. These properties make Miras valuable as standard candles for distance measurements, which offers insights into the structure of the Milky Way and other galaxies, thanks to their well-established period-luminosity relation \citep[e.g.,][]{10.1093/mnras/221.4.879, feast1989period, catchpole2016age}. In contrast, semi-regular variables (SRVs), another type of LPVs, typically show multiple pulsation modes in Fourier spectra. While this complicates luminosity estimation using period-luminosity relations, it provides more independent and comprehensive information about the fundamental properties of these stars. Moreover, since SRVs are far more numerous than Miras \citep{jayasinghe2021}, providing accurate luminosity estimates would enable a more detailed distance map of the sky. In this context, \cite{tabur2010period} employed the three strongest pulsation modes in Fourier spectra to calculate the joint probability density function (PDF) for deriving absolute magnitudes, while \cite{hey2023far} refined the precision of these estimates by incorporating pulsation amplitudes.

The studies by \cite{tabur2010period} and \citet{hey2023far} both focus on analysis in the frequency domain, which is often complicated by spectral window effects introduced by observational gaps. Two main methods have been adopted to address these artifacts. One involves directly handling gaps, either by filling them through interpolation \citep[e.g.,][]{pires_gap_2015} or by removing them entirely \citep[e.g.,][]{hekker_oscillation_2010}. 
These methods, however, become ineffective when gaps are substantial, as they may introduce offsets in the frequency measurements \citep{bedding2022dealing}. The other approach uses Gaussian Processes \citep[GP,][]{Rasmussen2004} to model the data in the time domain \citep{brewer2009gaussian}. For example, \citet{pereira2019gaussian} demonstrated that GP could accurately model granulation and oscillations in simulated \tess\ data, even in the presence of large data gaps. More recently, \citet{hey2024precise} demonstrated that GP-based models are robust against data gaps and cadence variations, and can significantly improve the precision of global seismic parameter measurements.

The objective of this work is to use Gaussian Processes to conduct asteroseismic analysis of the OGLE small amplitude red giants (OSARGs, \citealt{soszynski2004optical,2004MNRAS.349.1059W}) in the time domain using OGLE-II, OGLE-III \citep{soszynski2009optical, soszynski2013optical}, and OGLE-IV \citep{udalski_ogle-iv_2015} light curves that span approximately 26 years. This long-duration dataset provides an unprecedented level of frequency resolution for asteroseismic studies. Traditionally, OSARGs are classified as SRVs. Previous studies have shown that OSARGs exhibit solar-like oscillations both observationally \citep[e.g, ][]{2010A&A...524A..88D,10.1093/mnras/stt1685, yu2020asteroseismology} and theoretically \citep[e.g., ][]{takayama_pulsation_2013, 2020MNRAS.499.4687C}. More recently, \citet{trabucchi2024selfexcitedpulsationsinstabilitystrip} concluded that there is a transition in excitation mechanisms: OSARGs are driven by solar-like oscillations, while SRVs undergo self-excited pulsations. In light of these findings, our study aims to precisely measure the global seismic parameters typically associated with Sun-like oscillators, specifically the frequency of maximum power (\numax) and the large frequency separation (\Dnu), for stars in both the Large Magellanic Cloud (LMC) and the Galactic Bulge (BLG). Using the relation between these seismic parameters and \mk\ for stars in the LMC, we determine the \mk\ of BLG stars with high precision. The paper is organized as follows: Section~\ref{sec:target} describes the OGLE data and preliminary study; Section~\ref{sec:method} outlines the method we employ for extracting global seismic parameters; Section~\ref{sec:result} presents our main results; and the paper is closed with conclusions in Section~\ref{sec:conclusion}.

\section{Data and Sample selection} \label{sec:target}

\subsection{OGLE-IV Data}

\sloppy
The Optical Gravitational Lensing Experiment \citep[OGLE, ][]{1992AcA....42..253U} is a sky survey project conducted with the 1.3-meter Warsaw Telescope, located at the Las Campanas Observatory in Chile. Launched in 1992 (OGLE-I), the project is currently in its fourth phase \citep[OGLE-IV,][]{udalski_ogle-iv_2015}, which began in March 2010. This latest phase has significantly enhanced observational capabilities. The telescope now features a mosaic camera with 32 CCD detectors, each with a resolution of 2048×4102 pixels. It provides high-quality fields of view approximately 1.4 deg$^2$, covering over 3,000 square degree of the sky and regularly monitoring more than a billion sources, primarily in the Galactic center and the Magellanic Clouds. Furthermore, the project has consistently used filters that closely match the standard \textit{VI} bands, with $I$-band measurements achieving very high precision, typically around 0.005 mag for stars brighter than $I \lesssim 16$ mag.

We used $I$-band light curve data from the OGLE-II, OGLE-III \citep{soszynski2009optical, soszynski2013optical}, and OGLE-IV \citep{udalski_ogle-iv_2015} databases, which provide long-duration light curves with observations spanning from the year 1997 to 2023. We focused on the OGLE small amplitude red giants \citep[OSARGs,][]{soszynski2004optical, 2004MNRAS.349.1059W}, comprising 79,200 stars in the LMC and 192,643 in the BLG. For each star in the selected sample, the primary period (\p1), determined through an iterative fitting and subtraction process and further validated by visual inspection, is provided in the OGLE-III LPV catalog \citep{soszynski2009optical, soszynski2013optical} and ranges from 10 to 100 days. The classifications have inferred that this sample includes stars on the Red Giant Branch (RGB) or the Asymptotic Giant Branch (AGB) \citep{soszynski2009optical}.

\subsection{Sample Selection} \label{sec:pre_study}

We cross-matched our data with the 2MASS \citep{2006AJ....131.1163S} database and obtained 70,281 stars with high-quality photometric $K_\text{s}$-band data (\texttt{Qflg=`A’}). Adopting a $K_\text{s}$-band extinction value of 0.0372 mag from \citet{2012ApJ...753...71R} and a distance modulus of $18.477 \pm 0.004_\text{stat} \pm 0.026_\text{sys}$ mag for the LMC \citep{2019Natur.567..200P}, we derived the absolute $K_\text{s}$-band magnitude (\mk) for each star.

To refine the sample while preserving as many stars as possible, we first applied the Lomb-Scargle method to transform the time-domain data of LMC stars into the frequency domain. We then modeled the oscillation power excess using a Gaussian profile, with the background comprising a Harvey profile for granulation and a constant term for photon noise. The fitting was performed using a least-squares method with a maximum of 1,000 iterations. Stars with a signal-to-noise ratio (SNR) $<$ 1.5 were excluded, where the SNR is defined as the ratio of the peak of the fitted Gaussian profile to the background level. After removing 2,347 stars that failed to converge within the 1,000 iterations, and an additional 14,661 stars that did not meet the SNR criterion, a total of 53,273 unique targets were retained.

\begin{figure}[h]
    \centering
    \includegraphics[width=.47\textwidth]{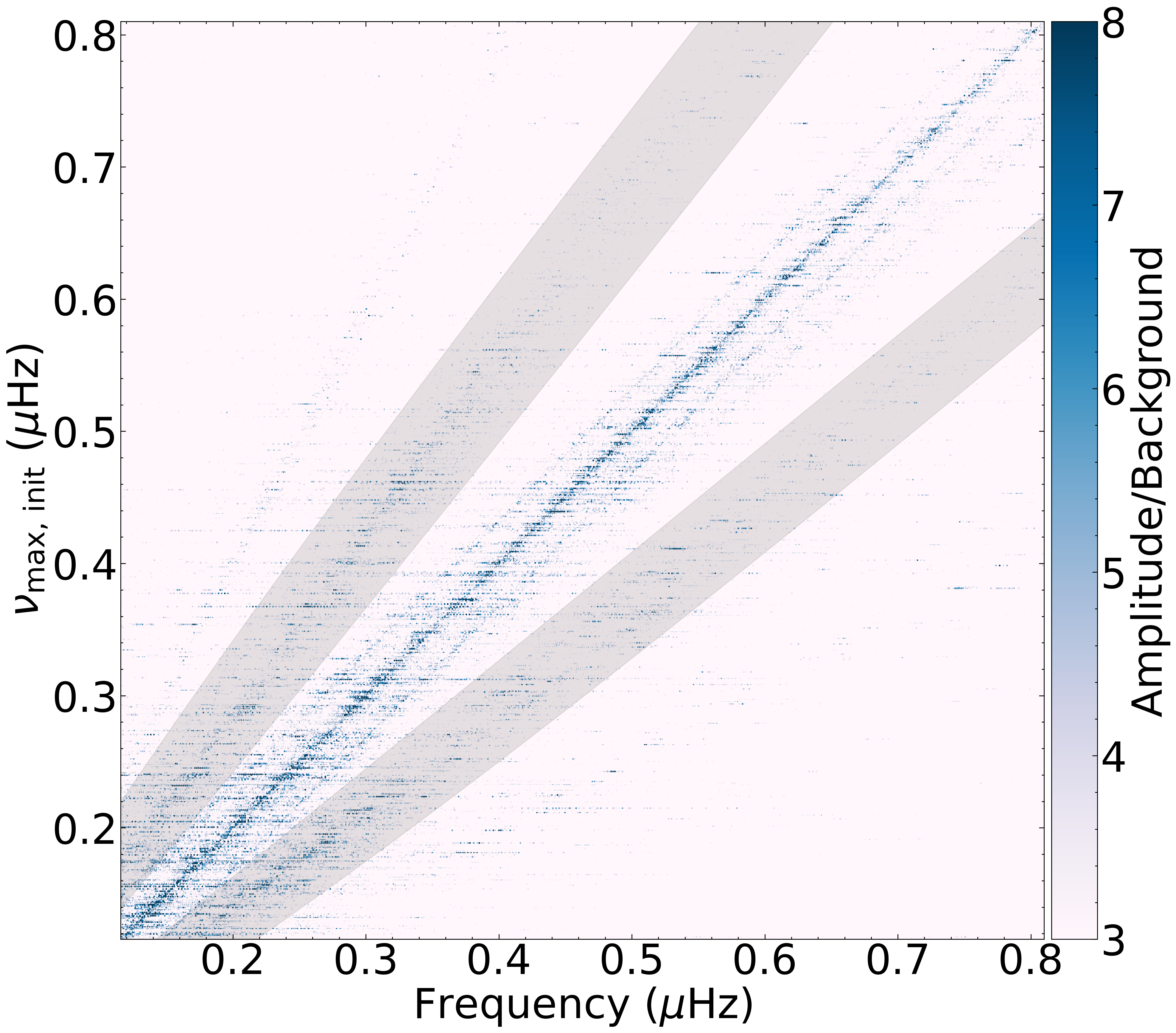}
    \caption{Stacked power spectra for 53,273 stars in the LMC, with granulation and photon noise (i.e. spectral background) removed from each spectrum. The y-axis label ``$\nu_\text{max, init}$'' corresponds to the primary period (\p1) expressed in microhertz ($\mu$Hz), obtained from the OGLE-III LPV catalog. The shaded region indicates the range defined for the sequence, with the enclosed data points used for further analysis (see text for details) The color bar indicates the amplitude of the background corrected spectrum (see text for details). The spectral window resulting from observation gaps (window function) is visible in the middle sequence.}
    \label{fig:stacks}
\end{figure}

To inspect the oscillation patterns of these stars, we divided their spectra by the background to obtain background-corrected power spectra (Amplitude/Background). These spectra were then stacked together, as shown in Figure~\ref{fig:stacks}. The spectra were sorted by the \p1, denoted as $\nu_{\text{max, init}}$, with higher values placed at the top. Figure~\ref{fig:stacks} reveals three distinct sequences, which likely correspond to three different radial orders ($n$). The sequence along the diagonal represents the primary mode, while evident frequency aliases, caused by the window function, appear as two lines on either side of the primary sequence.

To construct an appropriate prior relation, we processed the sequence on the stacked diagram. The detailed procedure is as follows:

\begin{enumerate}[label=\arabic*)]
    \item Extract the two side sequences adjacent to the main ridge in Figure~\ref{fig:stacks}. We manually selected the corresponding regions, as indicated by the shaded areas in Figure~\ref{fig:stacks}. The data points within these regions were then extracted. Each data point is described by three quantities: the frequency, the corresponding $\nu_{\text{max, init}}$, and the background-corrected amplitude represented by the colorbar.
    
    \item Define the quantity $\Delta\nu_{\text{init}}$. We calculated $\Delta\nu_{\text{init}}$ as the absolute difference between the frequency of each data point extracted in Step 1) and its corresponding $\nu_{\text{max, init}}$.
    
    \item Fit the relation between $\nu_{\text{max, init}}$ and $\Delta\nu_{\text{init}}$. We adopted a power-law form $\Delta\nu_{\text{init}} = a \nu_{\text{max, init}}^b$, and performed MCMC sampling to estimate the parameters. The background-corrected amplitudes were used as fitting weights. The best-fit parameters and their uncertainties were obtained from the median and standard deviation of the posterior distributions, and are summarized as follows:
    
\end{enumerate}

\begin{equation}
    \begin{cases}
        a = 0.248 \pm 0.001,\ b = 0.637 \pm 0.004, & \text{lower-right} \\
        a = 0.250 \pm 0.001,\ b = 0.828 \pm 0.004. & \text{upper-left}
    \end{cases}
\end{equation}

\noindent where the ``lower-right'' and ``upper-left'' sequences correspond to the lower-right and upper-left shaded regions in Figure~\ref{fig:stacks}, respectively. The final value of $\Delta \nu_{\rm init}$ was determined by averaging these best-fit parameter values, leading to:

\begin{equation}
    \Delta \nu_{\rm init} = 0.25\nu_\text{max,\ init}^{0.73},
\end{equation}

\noindent this relation was used to set priors for \numax\ and \Dnu\ in later analysis.

\section{Method} \label{sec:method}

A common characteristic of ground-based time-domain data is the presence of gaps, which introduce a window function in the Fourier spectrum and complicate the analysis of periodic signals.
 A straightforward approach to overcoming the issue is to fit the data in the time domain. Following \citet{brewer2009gaussian}, stellar oscillations can be modeled in the time domain as a series of damped harmonic oscillators using a Gaussian Process \citep[GP, ][]{Rasmussen2004} approach. By using appropriate parametrization, we can model global asteroseismological parameters such as \numax\ and \Dnu, as well as individual oscillation modes \citep{foreman2017fast}.

\subsection{Gaussian Process Theoretical Framework} \label{gp}

Gaussian process (GP) is a non-parametric Bayesian statistical method frequently utilized in the field of machine learning \citep{Rasmussen2004}. It is a type of stochastic process that assumes the value corresponding to each point (e.g., time and wavelength) is a normally distributed random variable. Furthermore, every finite collection of these random variables conforms to a multivariate Gaussian distribution. The distribution of a Gaussian process is the joint distribution of all these random variables, characterized by the covariance function and the mean function:

\begin{equation}
    f(x) \sim \mathcal{GP}(m(x), k(x, x')), 
\end{equation}

\noindent where $m(x)$ is mean function and $k(x,x')$ is covariance function or kernel. The kernel describes the similarity between any two points in the input space. Different kernel functions determine different prior distributions of functions. The mean function specifies our initial guess of the target function in the absence of any observational data.

\subsection{GP-based model for solar-like Oscillation} \label{sec:sho}

Following \citet{brewer2009gaussian}, solar-like oscillations can be regarded as a type of stochastic, damped harmonic oscillation that satisfies the following relation:

\begin{equation}
\left[ \frac{d^2}{dt^2} + \frac{\omega_0}{Q} \frac{d}{dt} + \omega_0^2 \right] y(t) = x(t), 
\end{equation}

\noindent where $\omega_0$ denotes the angular frequency of the undamped oscillation, $Q$ represents the quality factor, which is the ratio of the damping time to the oscillation period, and $x(t)$  describes a stochastic driving force. The corresponding power spectral density (PSD) is:

\begin{equation}
S(\omega) = \sqrt{\frac{2}{\pi}} \frac{S_0 \omega_0^4}{(\omega^2 - \omega_0^2)^2 + \omega_0^2 \omega^2 / Q^2}, \label{eq:oscillation}
\end{equation}

\noindent where $S_0$ is proportional to the spectral power at $\omega = \omega_0$ ($S(\omega_0) = \sqrt{2/\pi}S_0Q^2$). It is worth noting that when $Q = 1/\sqrt{2}$, the corresponding PSD often matches the functional form of model granulation in the frequency domain \citep{1985ESASP.235..199H}:

\begin{equation}
S(\omega) = \sqrt{\frac{2}{\pi}} \frac{S_0}{\left(\frac{\omega}{\omega_0}\right)^4 + 1}. \label{eq:granulation}
\end{equation}

The discussion of other $Q$ values and their description of oscillations can be found in \citet{foreman2017fast}.

\subsection{Model Construction}

We used the Simple Harmonic Oscillator (SHO) term in the Python library CELERITE2 \citep{foreman2017fast,celerite2} as the kernel for GP modeling. The corresponding functional form is:

\begin{equation}
\begin{aligned}
    &\hspace{2em} k(\tau; S_0, Q, \omega_0) = S_0 \omega_0 Q e^{\frac{\omega_0 \tau}{2Q}} \\
    & \quad \times
    \begin{cases}
        \cosh(\eta \omega_0 \tau) + \frac{1}{2\eta Q} \sinh(\eta \omega_0 \tau), & 0 < Q < \frac{1}{2}, \\
        2(1 + \omega_0 \tau), & Q = \frac{1}{2}, \\
        \cos(\eta \omega_0 \tau) + \frac{1}{2\eta Q} \sin(\eta \omega_0 \tau), & \frac{1}{2} < Q,
    \end{cases}
\end{aligned}
\end{equation}

\noindent where $\quad \eta = \left| 1 - (4Q^2)^{-1} \right|^{1/2}$. The kernel PSD follows the form given in Equation~\eqref{eq:oscillation}.

\subsubsection{\tt\string numax\_model} \label{sec:numax_model}

We first constructed a model to determine the \numax\ value and refer to the model as \mnumax. We used a combination of an overdamped SHO term with  $Q=1/\sqrt{2}$ to model granulation and a regular SHO term to model oscillations. Additionally, we added a small constant value ($\sigma$) to the diagonal elements of the covariance matrix to represent white noise. The equation for the granulation component can be related to the work of \citet{kallinger2014connection}:

\begin{equation}
S(\nu) = \frac{2\sqrt{2}}{\pi} \frac{a^2_{\text{gran}}/b_{\text{gran}}}{\left( \nu / b_{\text{gran}} \right)^4 + 1},
\end{equation}

\noindent where $a_{\rm gran}$ and $b_{\rm gran}$ are the characteristic amplitude and characteristic frequency of granulation respectively. For the oscillation component, we assumed the amplitude of oscillations follows a Gaussian distribution centered as \numax, which can be described by the following equation \citep{kallinger2014connection}:

\begin{equation}
    S(\nu) = H_{\rm osc} \exp \left( -\frac{(\nu - \nu_{\text{max}})^2}{2\sigma_\text{osc}^2} \right),
\end{equation}

\noindent where $H_{\rm osc}$ represents the oscillation height and $\sigma_\text{osc}$ denotes its width.

Compared to the original form of the kernel’s PSD (Equation~\eqref{eq:oscillation} and Equation~\eqref{eq:granulation}), the frequencies in the equations for both the granulation and oscillation components have been converted from angular frequencies to linear frequencies. A detailed description of the conversion between parameters can be found in \citet{pereira2019gaussian}. The priors for the input parameters of the \mnumax\ were shown in Table~\ref {tab:numax_model_prior}. We constrained the prior of \numax\ within the range ($\nu_{\rm max,\ init} - 1.5 \Delta \nu_{\rm init}, \, \nu_{\rm max,\ init} + 1.5 \Delta \nu_{\rm init}$), as the number of detectable modes for a single star is typically only 2 to 3, causing the position of \numax\ to be very close to the $\nu_{\rm max,\ init}$.

\begin{table}[h!] 
    \centering
    \begin{threeparttable}
        \caption{Input parameters and their priors of \mnumax.}
        \label{tab:numax_model_prior}
        
        \begin{tabularx}{.47\textwidth}{l@{\hskip 1.5cm}r}
            \toprule
            \toprule
            Parameter & Prior \\
            \midrule
            $\ln\left(a_{\rm gran}/\text{ppm}\right)$ & $\mathcal{U}(0,50)$ \\
            $b_{\rm gran}/\mu\rm{Hz}$ & $\mathcal{U}(0, 0.2\nu_{\rm max,\ init})$ \\
            $\ln\left(H_{\rm osc}/{\rm{ppm}}^2\mu\rm{Hz}^{-1}\right)$ & $\mathcal{U}(0,50)$ \\
            $\ln\left(Q_{\rm osc}\right)$ & $\mathcal{U}(0,10)$ \\
            $\nu_{\rm max}/\mu\rm{Hz}$ & $\mathcal{N}(\nu_{\rm max,\ init},\ 0.5\Delta\nu_{\rm init})$ \\
            $\sigma/\rm{ppm}$ & $\mathcal{N}(0,1)$ \\
            \bottomrule
        \end{tabularx}
        
        \begin{tablenotes}[flushleft]
            \item \textbf{Note.} The subscripts ``gran'' and ``osc'' represent granulation and oscillation, respectively. $\mathcal{U}$ and $\mathcal{N}$ denote the uniform and normal distributions, respectively.
        \end{tablenotes}
    \end{threeparttable}
\end{table}

\subsubsection{\tt\string dnu\_model} 

As described in Section~\ref{sec:target}, the three sequences in Figure~\ref{fig:stacks} likely represent different oscillation modes with distinct radial orders ($n$), suggesting that multiple modes can be excited in most stars, although these modes may not be prominent in the power spectrum of an individual star. We hence construct another model to determine the \Dnu{} values and refer to the model as \mdnu. We used an overdamped SHO term to describe the granulation background and three regular SHO terms to fit three individual oscillation modes around \numax. This model is based on Equations (57) and (58) from \citet{foreman2017fast}, but since the value of \numax\ for our sample is very low ($< 1\ \mu\rm{Hz}$), we removed the parameter $\epsilon$, which is used to shifts the location of oscillations on the power spectrum, as it would introduce degeneracy with \numax. We also varied the amplitude of each mode by adjusting the parameter $A$. Therefore, in our case, the functions that describe the angular frequency of oscillation modes are:

\begin{equation}
\omega_{0,j} = 2 \pi (\nu_{\text{middle}} + j \Delta \nu), \label{eq:w}
\end{equation}

\noindent where subscript ``middle'' corresponds to the central mode in Figure~\ref{fig:stacks}, i.e., the frequency of the central peak. The value of $j$ can be 0, -1, or +1, with -1 and +1 corresponding to the two modes surrounding the central mode. The functions that describe the amplitude of oscillation modes are:

\begin{equation}
S_{0,j} = \frac{A_j}{Q^2} \exp\left( -\frac{[j \Delta \nu]^2}{2 W^2} \right), 
\end{equation}

\noindent where $A_j$, $W$, $Q$ are nuisance parameters. The priors for the input parameters of the \mdnu\ were shown in Tabel~\ref{tab:mode_model_prior}. The priors for $\nu_{\rm middle}$ were constrained to the same range as \numax\ in the \mnumax.

\begin{table}[h!] 
    \centering
    \begin{threeparttable}
        \caption{Input parameters and their priors of \mdnu. }
        \label{tab:mode_model_prior}
        \begin{tabularx}{0.47\textwidth}{l@{\hskip 2.5cm}r}
            \toprule
            \toprule
            Parameter & Prior \\
            \midrule
            $\text{ln}(S_{\rm 0,gran}/{\text{ppm}}^2\text{day}$) & $\mathcal{U}(5,50)$ \\
            $\omega_{\rm gran}/\rm day^{-1}$ & $\mathcal{U}(0,0.1)$ \\
            $Q_{\rm gran}$ & $\mathcal{B}(2,2)$ \\
            $\text{ln}(A_j/{\text{ppm}}^2\text{day}$) & $\mathcal{U}(5,50)$ \\
            $W/\rm day^{-1}$ & $\mathcal{N}(0,1)$ \\
            $\text{ln}(Q_{\rm osc})$ & $\mathcal{U}(0,10))$ \\
            $\nu_{\rm middle}/\rm \mu Hz$ & $\mathcal{N}\rm (\nu_{\rm max,\ init}, \Delta \nu_{\rm init})$ \\
            $\Delta \nu/\rm \mu Hz$ & $\mathcal{N}(\rm \Delta \nu_{\rm init}, 0.3\Delta\nu_{\rm init})$ \\
            $\sigma/\text{ppm} $ & $\mathcal{N}(0,1)$ \\
            \bottomrule
        \end{tabularx}

        \begin{tablenotes}[flushleft]
            \item \textbf{Note.} $\mathcal{B}$ denotes the beta distribution. The definition of $\sigma$ is consistent with that used in the \mnumax.
        \end{tablenotes}
                
    \end{threeparttable}

\end{table}

\subsection{Fitting Scheme} \label{sec:regression}

We used a Bayesian approach to perform Gaussian process regression, with the log-likelihood function given by:

\begin{equation}
\ln (X) = -\frac{1}{2} \mathbf{r}^T \mathbf{K}^{-1} \mathbf{r} - \frac{1}{2} \ln |\det \mathbf{K}| - \frac{N}{2} \ln 2\pi,
\end{equation}

\noindent where $X$ represents the set of parameters used to build the kernel, \(\mathbf{r}\) represents the vector of residuals obtained after subtracting the mean function, $N$ denotes the number of data points, and \(\mathbf{K}\) is the covariance matrix with elements defined as

\begin{equation}
K_{ij}(X) = \sigma_i^2 \delta_{ij} + k(\tau_{i,j}; X),
\end{equation}

\noindent for each pair of points indexed by \(i\) and \(j\), \(\sigma_i\) represents the uncertainty in the observation of the \(i\)-th point, $\delta_{ij}$ is the Kronecker notation, and $\tau_{i,j}=|t_i - t_j|$ is the absolute distance in time between points $i$ and $j$, $k$ is a kernel function parameterized by $X$. In this work, the mean function was set as a small constant, which varies between stars and is treated as a variable during sampling.

We used the No-U-Turn Sampler (NUTS) in the \text{P\scalebox{0.8}{Y}MC3} \citep{salvatier2016probabilistic} to sample in the parameter space. For each star, we ran two chains simultaneously. We randomly selected 2,000 stars for testing and found that 1,000 tuning steps and 1,000 draw steps were sufficient for parameter convergence. We used the Gelman-Rubin diagnostic \citep[$\hat{R}$,][]{brooks1998general} to assess parameter convergence. When $\hat{R}$ $\approx$ 1, indicating that the distributions of this parameter across different chains are highly similar, we considered the parameter to have converged. In addition, we excluded stars that exhibit more than 200 divergences after running the model, as their parameter posterior distributions were not considered reliable.

For LMC stars, we first applied \mnumax\ to estimate \numax\ and excluded stars with a signal-to-noise ratio (S/N) below 5, where the definition of S/N follows that given in Section~\ref{sec:pre_study}. After this step, the number of remaining stars was reduced from 53,273 to 16,489. Subsequently, we applied \mdnu\ to this subset to determine \Dnu. The initial position of each parameter comes from the maximum a posteriori (MAP) estimation. For BLG stars, given their large number (192,643 in total), we initially fitted the power spectrum using the method described in Section~\ref{sec:target}, selected stars with S/N $>$ 3, and then applied the same procedure used for the LMC stars. The purpose of this process is to conserve resources. In the end, the number of remaining BLG stars was reduced from 192,643 to 23,266.

\begin{figure}[ht]
    \centering
    \includegraphics[width=1.\linewidth]{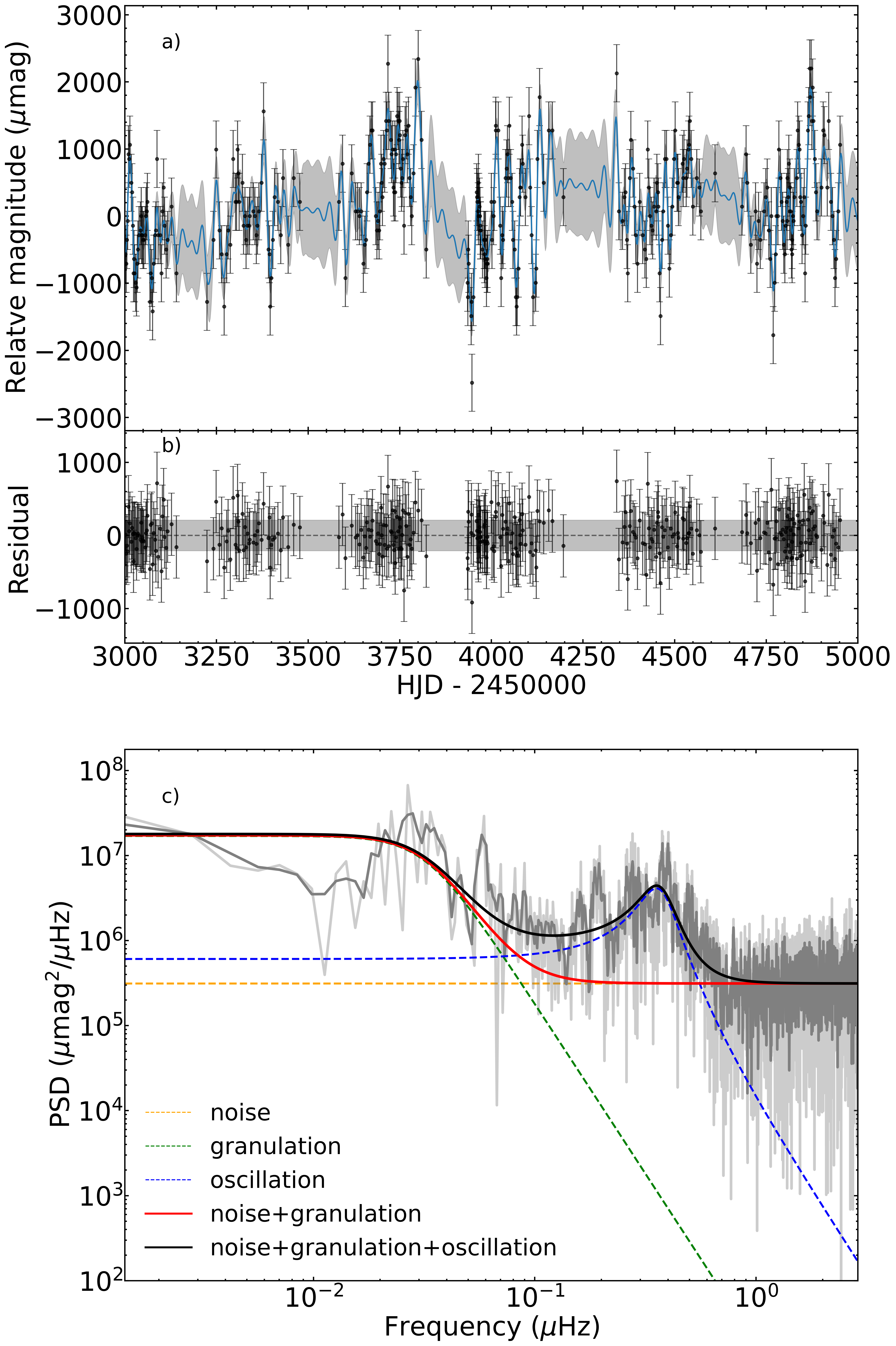}
    \caption{\textbf{panel a)}: Light curve fitting results of \texttt{OGLE-LMC-LPV-23573} using the \mnumax. The thin blue line represents
    the best fit, and the gray area indicates the 68\% confidence interval (1$\sigma$). \textbf{panel b)}: The residuals of the light curve fit for \mnumax, the dashed line marks residual=0, with the gray area representing the same confidence interval as in the upper panel. Note that the full range is from 2167.8 to 10409.7, but only the interval from 3,000 to 5,000 is shown in the plot. \textbf{panel c)}: Representative power spectral density showing the best-fitting \mnumax\ models for noise, granulation, and oscillations, along with their combinations, all performed in the time domain (see the legend for individual model components). The light gray curve represents the original power spectral density calculated using the Lomb-Scargle algorithm, while the dark gray line illustrates the smoothed spectrum, averaged over a width of 0.00145 $\mu$Hz.} 
    \label{fig:numax_model_res}
\end{figure}

\begin{figure}[ht]
    \centering
    \includegraphics[width=1.\linewidth]{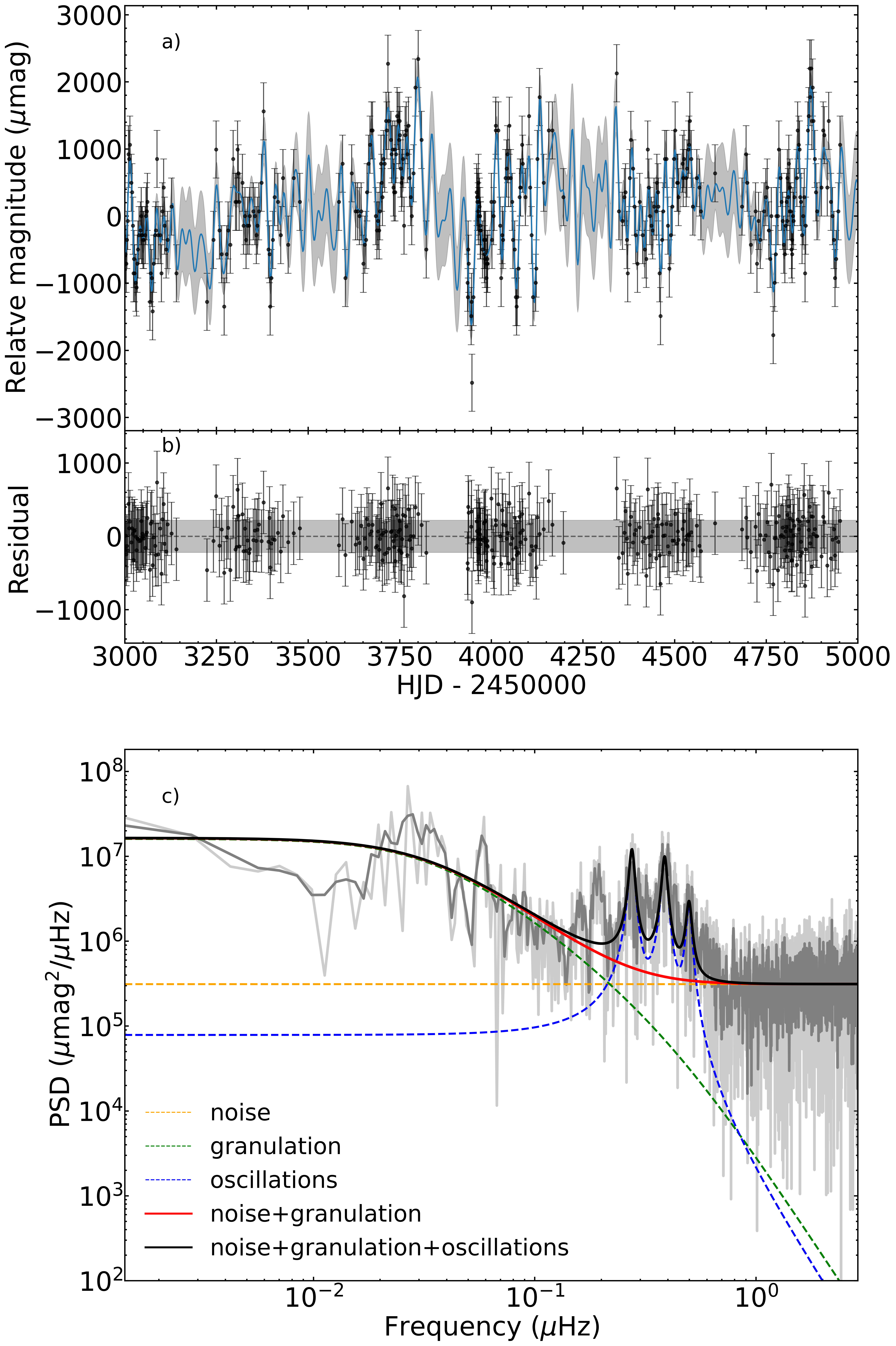}
    \caption{Similar to Figure~\ref{fig:numax_model_res}, but now replacing the Gaussian envelope characterizing the oscillation power excess with our \mdnu, which models three modes included in the power excess.}
    \label{fig:dnu_model_res}
\end{figure}

Taking the star {\tt\string OGLE-LMC-LPV-23573} as an example, Figures~\ref{fig:numax_model_res} and \ref{fig:dnu_model_res} present the direct light curve fitting results using the \mnumax\ and \mdnu, along with their corresponding representations in the frequency domain. Further examples can be found in the Appendix~\ref{appendixA}.

\section{Results} \label{sec:result}

\begin{figure}[ht]
    \centering
    \includegraphics[width=1.\linewidth]{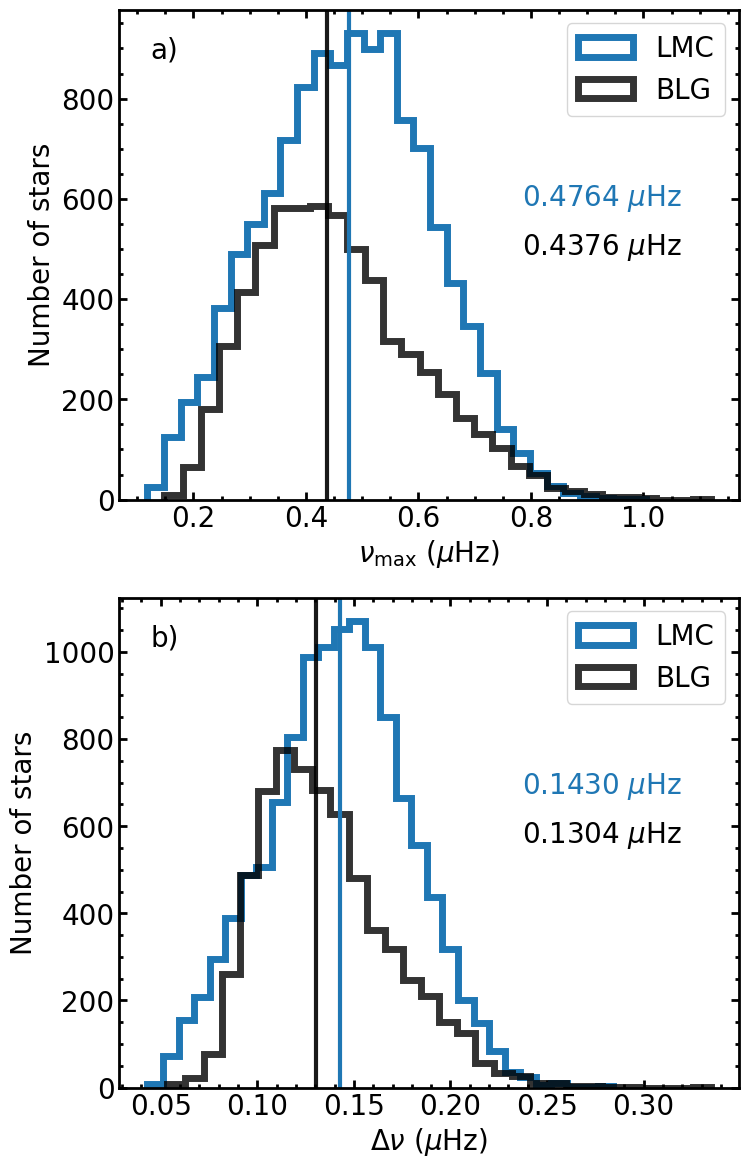}
    \caption{The distribution of the \numax\ (\textbf{panel a}) and \Dnu\ (\textbf{panel b}) for 12,055 LMC stars (blue) and 6,391 BLG stars (black). The vertical lines represent the median values, as denoted in the plots.}
    \label{fig:numax_dnu_dis}
\end{figure}

\begin{figure*}[ht]
    \centering
    \includegraphics[width=1.\linewidth]{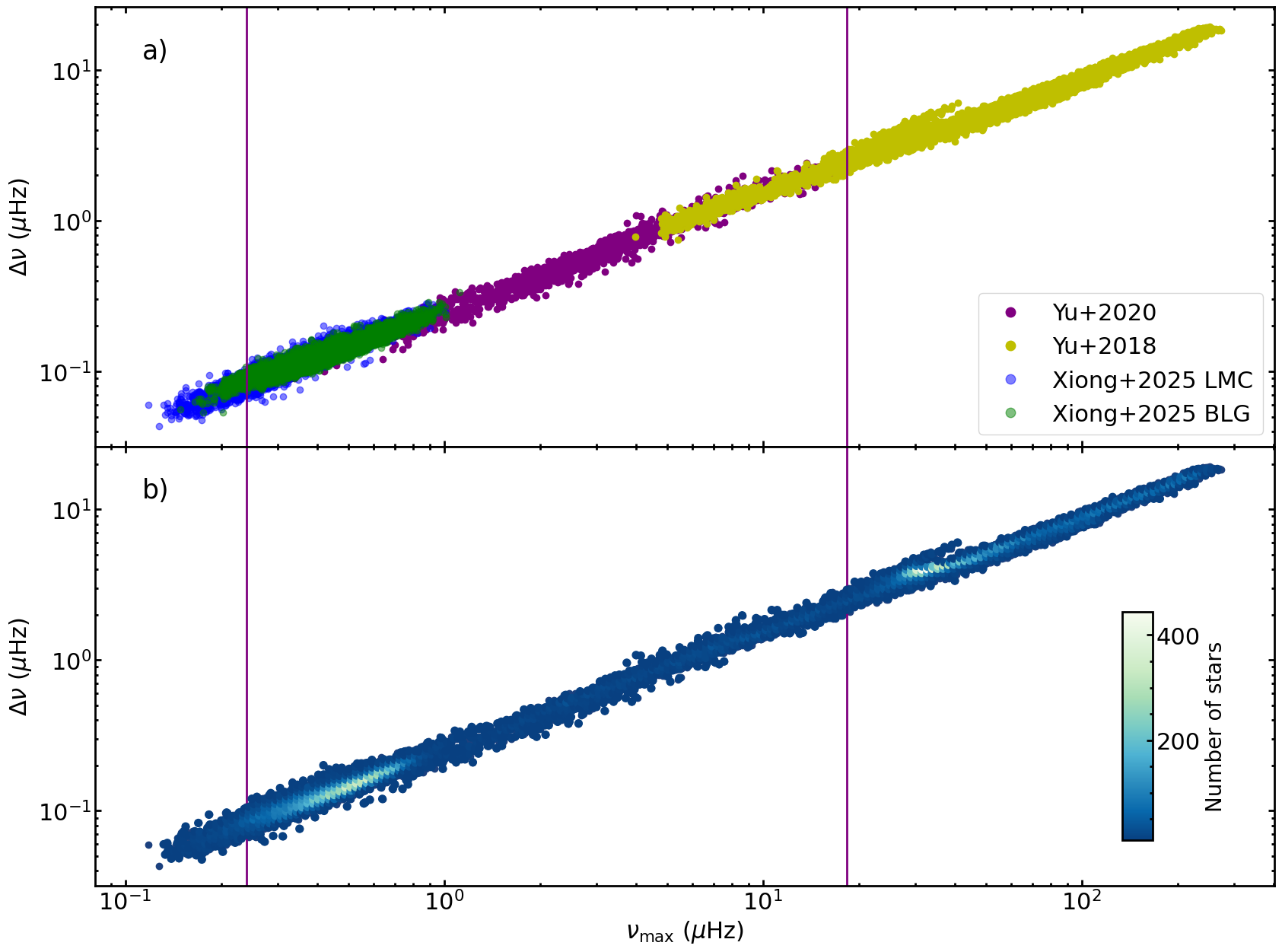}
    \caption{\textbf{Panel a)}: \numax-\Dnu\ diagram for LPVs in the LMC (blue) and BLG (green) using OGLE data and for lower-luminosity (yellow, \citealt{yu2018asteroseismology}) and higher-luminosity (purple, \citealt{yu2020asteroseismology}) red giants observed by \kepler. \textbf{Panel b)}: Similar to \textbf{Panel a)} but now color-coded by star number density. The purple vertical lines in both panels mark the upper and lower limits of \numax\ for \kepler\ higher-luminosity red giants.}
    \label{fig:numax_dnu}
\end{figure*}

After the analysis using \mnumax\ and \mdnu, we identified 12,055 stars in the LMC and 6,391 in the BLG with measurements of \numax\ and \Dnu, verified through visual inspection. The distribution of these values is illustrated in Figure~\ref{fig:numax_dnu_dis}, which clearly indicates that most stars in our sample have values of \numax\ below 1 $\mu$Hz. The \numax-\Dnu\ relation measured in this work for the LMC and BLG samples are in good agreement with the \textit{Kepler} red giants \citep{yu2018asteroseismology, yu2020asteroseismology}, as shown in Figure~\ref{fig:numax_dnu}, suggesting that OSARGs exhibit similar oscillation features and are likely solar-like oscillators, consistent with previous studies \citep[e.g.,][]{2010A&A...524A..88D,takayama_pulsation_2013,yu2020asteroseismology,trabucchi2024selfexcitedpulsationsinstabilitystrip}. Detailed results for individual stars in the LMC can be found in Table~\ref{tab:starlist_lmc}, while the results for the BLG are presented in Table~\ref{tab:starlist_blg}.

We focused on LMC stars to explore the relation between seismic parameters and the measured absolute magnitude \mk. Based on this relation, we calculated the \mk\ for the BLG sample, and by combining this with extinction (\citealt{2012A&A...543A..13G, 2013ApJ...769...88N, 2014A&A...566A.120S, 2016ApJ...818..130B}), we derived the distance to each star in the BLG.

\begin{table}[!htbp]
    \centering
    \renewcommand{\arraystretch}{1.4}
    \begin{threeparttable}
        \caption{Measured \numax, \Dnu\ and calculated \mk\ in this work for each star in the LMC.}
        \label{tab:starlist_lmc}
        \begin{tabularx}{.47\textwidth}{@{\hskip .0cm} c @{\hskip .35cm} c @{\hskip .35cm} c @{\hskip .35cm} c @{\hskip .35cm} c @{\hskip .35cm} c @{\hskip .35cm} c @{\hskip .35cm} c @{\hskip .35cm} c}
        \toprule
        \toprule
    
        \shortstack{ID\\ \;} & \shortstack{\numax\ \\ ($\mu$Hz)} & \shortstack{$\sigma_{\nu_{\rm max}}^\text{68th}$ \\ ($\mu$Hz)} & \shortstack{\Dnu\ \\ ($\mu$Hz)} & \shortstack{$\sigma_{\Delta \nu}^\text{68th}$ \\ ($\mu$Hz)} & \shortstack{$M_{K}$ \\ (mag)} & \shortstack{$\sigma_{M_{K}}$ \\ (mag)} \\
    
        \midrule
        19341 & 0.5180 & 0.0184 & 0.1706 & 0.0066 & -5.7014 & 0.0397 \\
        19503 & 0.4973 & 0.0179 & 0.1473 & 0.0042 & -6.0154 & 0.0420 \\
        30672 & 0.4175 & 0.0180 & 0.1373 & 0.0086 & -5.7094 & 0.0420 \\
        42334 & 0.4047 & 0.0193 & 0.1254 & 0.0069 & -5.7594 & 0.0420 \\
        52203 & 0.4322 & 0.0159 & 0.1300 & 0.0024 & -6.1194 & 0.0397 \\
        61343 & 0.3843 & 0.0106 & 0.1191 & 0.0050 & -6.4464 & 0.0375 \\
        64420 & 0.8022 & 0.0196 & 0.2236 & 0.0079 & -5.2094 & 0.0737 \\
        65985 & 0.5961 & 0.0238 & 0.1592 & 0.0058 & -5.9884 & 0.0389 \\
        75631 & 0.4707 & 0.0192 & 0.1353 & 0.0108 & -5.8264 & 0.0375 \\
        84354 & 0.5129 & 0.0197 & 0.1418 & 0.0052 & -6.1214 & 0.0420 \\
        \bottomrule
        \end{tabularx}
        \begin{tablenotes}[flushleft]
            \item \textbf{Note.} The prefix “OGLE-LMC-LPV-” for IDs has been omitted. The full table includes 12,055 stars; here, ten randomly selected stars are shown (sorted by star ID). $\sigma^{\text{68th}}$ represents the uncertainty of the posterior distribution at the 68\% confidence level. The \mk\ values here are derived using the method described in Section~\ref{sec:pre_study}. The complete dataset is available online.
        \end{tablenotes}
    \end{threeparttable}
\end{table}

\begin{figure*}[!htbp]
    \centering
    \includegraphics[width=1.\linewidth]{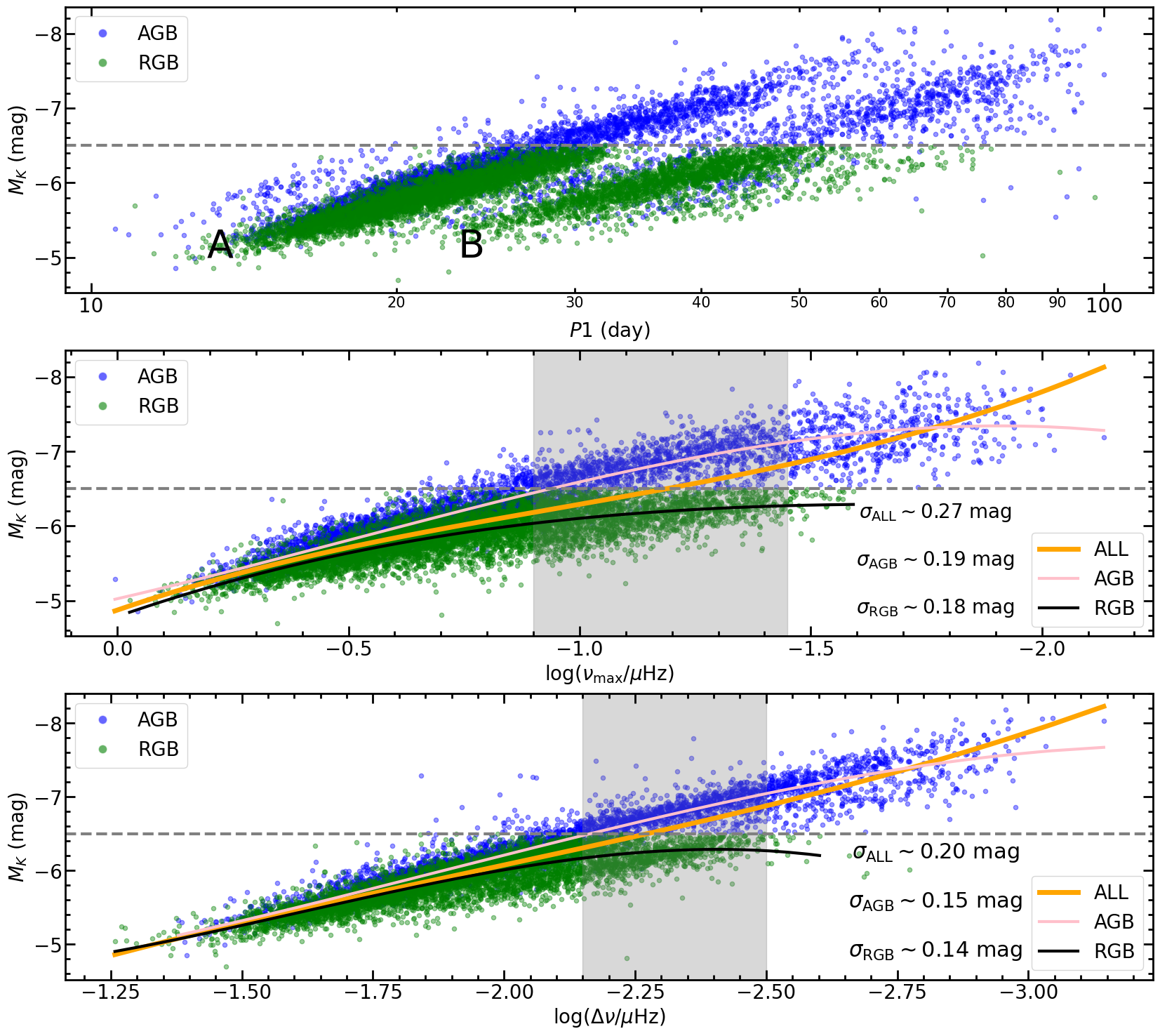}
    \caption{Absolute magnitude (\mk) versus the primary period (\p1, top panel), \numax\ (middle panel), and \Dnu\ (lower panel) for 12,055 OSARGs in the LMC. The values of \numax\ and \Dnu\ have been log-transformed to achieve a better fit. The upper panel shows the two well-known sequences, A and B, which merge when plotted as a function of \numax\ and form a tighter relation for \Dnu\. The gray horizontal dashed line marks the \mk\ value at the tip of the RGB. In the lower two panels, the black, red, and yellow lines represent the best-fitting third-order polynomials for RGB, AGB, and all the stars, respectively, with the scatter between the models and observations indicated in the plot. The gray-shaded regions highlight the significantly reduced scatter, suggesting higher precision in \mk\ estimates derived from \numax\ or \Dnu\ compared with \p1.}
    \label{fig:nx_dn_mk_fit}
\end{figure*}

\begin{table*}[!htbp]
    \centering
    \renewcommand{\arraystretch}{1.4}
    \begin{threeparttable}
        \caption{Measured \numax, \Dnu, calculated \mk\ values, and distance modulus for each BLG star.}
        \label{tab:starlist_blg}
        
        \begin{tabularx}{\textwidth}{@{\hskip .9cm} c @{\hskip .8cm} c @{\hskip .8cm} c @{\hskip .8cm} c @{\hskip .8cm} c @{\hskip .8cm} c @{\hskip .8cm} c @{\hskip .8cm} c @{\hskip .8cm} c @{\hskip .8cm} c }
        \toprule
        \toprule
        \shortstack{ID\\ \;} & \shortstack{\numax\ \\ ($\mu$Hz)} & \shortstack{$\sigma_{\nu_{\rm max}}^\text{68th}$ \\ ($\mu$Hz)} & \shortstack{\Dnu\ \\ ($\mu$Hz)} & \shortstack{$\sigma_{\Delta \nu}^\text{68th}$ \\ ($\mu$Hz)} & \shortstack{\text{\mk} \\ (mag)} & \shortstack{$\sigma_{M_{K}}$ \\ (mag)} & \shortstack{$\mathrm{\mu}$ \\ (mag)} & \shortstack{$\sigma_\mathrm{\mu}$ \\ (mag)} & \shortstack{\text{Flag}\\ \;} \\
        
        \midrule

        006523 & 0.2354 & 0.0139 & 0.0934 & 0.0057 & NaN & NaN & NaN & NaN & 1 \\
        007457 & 0.2130 & 0.0089 & 0.0885 & 0.0126 & -6.98 & 0.26 & 15.44 & 0.25 & 0 \\
        027074 & 0.1889 & 0.0095 & 0.0654 & 0.0056 & -7.26 & 0.24 & 15.60 & 0.26 & 0 \\
        029373 & 0.2973 & 0.0132 & 0.0908 & 0.0025 & NaN & NaN & NaN & NaN & 1 \\
        050964 & 0.8339 & 0.0260 & 0.2071 & 0.0088 & NaN & NaN & NaN & NaN & 2 \\
        071016 & 0.3071 & 0.0085 & 0.1225 & 0.0103 & -6.24 & 0.19 & 14.77 & 0.28 & 0 \\
        103342 & 0.5659 & 0.0153 & 0.1459 & 0.0026 & -5.91 & 0.15 & 14.53 & 0.28 & 0 \\
        136795 & 0.7547 & 0.0234 & 0.2084 & 0.0143 & -5.39 & 0.14 & NaN & NaN & 0, 3, 4 \\
        174489 & 0.3120 & 0.0108 & 0.1119 & 0.0081 & -6.24 & 0.29 & 15.21 & 0.25 & 0 \\
        184205 & 0.4877 & 0.0162 & 0.1279 & 0.0031 & -6.10 & 0.18 & 14.60 & 0.27 & 0 \\

        \bottomrule
        \end{tabularx}
        \begin{tablenotes}[flushleft]
        \item \textbf{Note.} Similar to Tabel~\ref{tab:starlist_lmc}, but for stars in the BLG. The prefix ``OGLE-BLG-LPV-'' for IDs has been omitted.  $\mathrm{\mu}$ denotes the distance modulus. The uncertainty in the distance modulus ($\sigma_{\mu}$) is derived from Monte Carlo analysis, which incorporates the uncertainties in the apparent magnitude, the absolute magnitude derived in this work, and extinction (see text for more details). The values 0, 1, and 2 in the Flag column indicate that the joint PDF is unimodal, bimodal, and unreliable, respectively, while 3 and 4 indicate the poor photometric quality of the $K_\text{s}$-band magnitudes from 2MASS (\texttt{Qflg$\neq$`A')} and unreliable extinction values ($A_{K_\text{s}}$) derived based on the OGLE extinction map (\texttt{QF$\neq$0)}, respectively. The complete table contains 6,391 stars and is available online.
        \end{tablenotes}
    
    \end{threeparttable}
    
\end{table*}

\subsection{\numax-\mk\ and \Dnu-\mk\ relations}

In Figure~\ref{fig:nx_dn_mk_fit}, we compared three luminosity indicators: the primary period ($P$1), the frequency at maximum power (\numax), and the frequency spacing (\Dnu) for the LMC sample. The $P$1-$M_{\rm K}$ diagram in the top panel reveals two sequences, labeled A and B according to the nomenclature established by \citet{wood1999proc}.

Compared to \p1, both \numax\ and \Dnu\ exhibit significantly reduced scatter as a function of \mk, as shown in the middle and bottom panels in Figure~\ref{fig:nx_dn_mk_fit}. Additionally, no distinct sequences were identified for \numax\ or \Dnu; instead, a bimodal distribution was observed within the mixed range of RGB and AGB stars (highlighted by the gray shading). However, it is important to note that only a limited number of stars have data of sufficient quality to allow deriving \numax\ and \Dnu\ using the GP model in this work. The requirement of data quality limits the applicability of this method.

We fitted our results using third-order polynomial functions for the entire star sample, as well as for the RGB and AGB samples in a logarithmic space. The scatter, quantified as the root mean square error (RMSE), is 0.27 mag for the \numax-\mk\ relation and 0.21 mag for the \Dnu-\mk\ relation when considering the whole sample. This indicated that \Dnu\ serves as a more precise luminosity indicator compared to \numax. Furthermore, we observed that the scatter could be significantly reduced when the classifications of RGB or AGB are taken into account (refer to $\sigma_{\rm RGB}$ and $\sigma_{\rm AGB}$ in Figure~\ref{fig:nx_dn_mk_fit}). The fitted parameters were summarized in Table~\ref{tab:fit_parameters}.

\begin{table}[htbp]
    \centering
    \renewcommand{\arraystretch}{1.6}
    \begin{threeparttable}
        \caption{The Fitting results in Figure~\ref{fig:nx_dn_mk_fit}.}
        \label{tab:fit_parameters}
        \begin{tabularx}{0.47\textwidth}{l@{\hskip .2cm}c@{\hskip .5cm}c@{\hskip .5cm}r}
        \toprule
        \toprule

        \text{Sample} & \(\displaystyle \log\left(\frac{\nu_{\text{max}}}{\mu\text{Hz}}\right)\) &  $\displaystyle \log\left(\frac{\Delta \nu}{\mu \text{Hz}}\right)$ & \textbf{P*} \\
        
        \midrule
        \multirow{3}{*}{\text{ALL}} 
        & $0.376 \pm 0.034$ & $0.242 \pm 0.041$ & \textbf{a3} \\
        & $1.085 \pm 0.100$ & $1.423 \pm 0.262$ & \textbf{a2} \\
        & $2.127 \pm 0.092$ & $4.312 \pm 0.551$ & \textbf{a1} \\
        & $-4.873 \pm 0.025$ & $-1.206 \pm 0.381$ & \textbf{a0} \\
    
        \midrule
        \multirow{3}{*}{\text{AGB}} 
        & $-0.273 \pm 0.032$ & $-0.305 \pm 0.040$ & \textbf{a3} \\
        & $-0.406 \pm 0.100$ & $-1.723 \pm 0.276$ & \textbf{a2} \\
        & $1.433 \pm 0.095$ & $-1.440 \pm 0.575$ & \textbf{a1} \\
        & $-5.025 \pm 0.027$ & $-4.626 \pm 0.423$ & \textbf{a0} \\
        \midrule
        \multirow{3}{*}{\text{RGB}} 
        & $0.131 \pm 0.065$ & $-0.846 \pm 0.093$ & \textbf{a3} \\
        & $0.968 \pm 0.160$ & $-4.086 \pm 0.547$ & \textbf{a2} \\
        & $2.153 \pm 0.123$ & $-4.974 \pm 1.065$ & \textbf{a1} \\
        & $-4.787 \pm 0.029$ & $-6.376 \pm 0.686$ & \textbf{a0} \\
        \bottomrule
    \end{tabularx}

    \begin{tablenotes}[flushleft]
    \item \textbf{Note.} \textbf{a3}, \textbf{a2}, \textbf{a1}, and \textbf{a0} represent the coefficients of the 3rd, 2nd, 1st and 0th order terms of the quadratic polynomial, respectively.

    \end{tablenotes}
    \end{threeparttable}
\end{table}

\begin{figure*}[htbp]
    \centering
    \includegraphics[width=1.\linewidth]{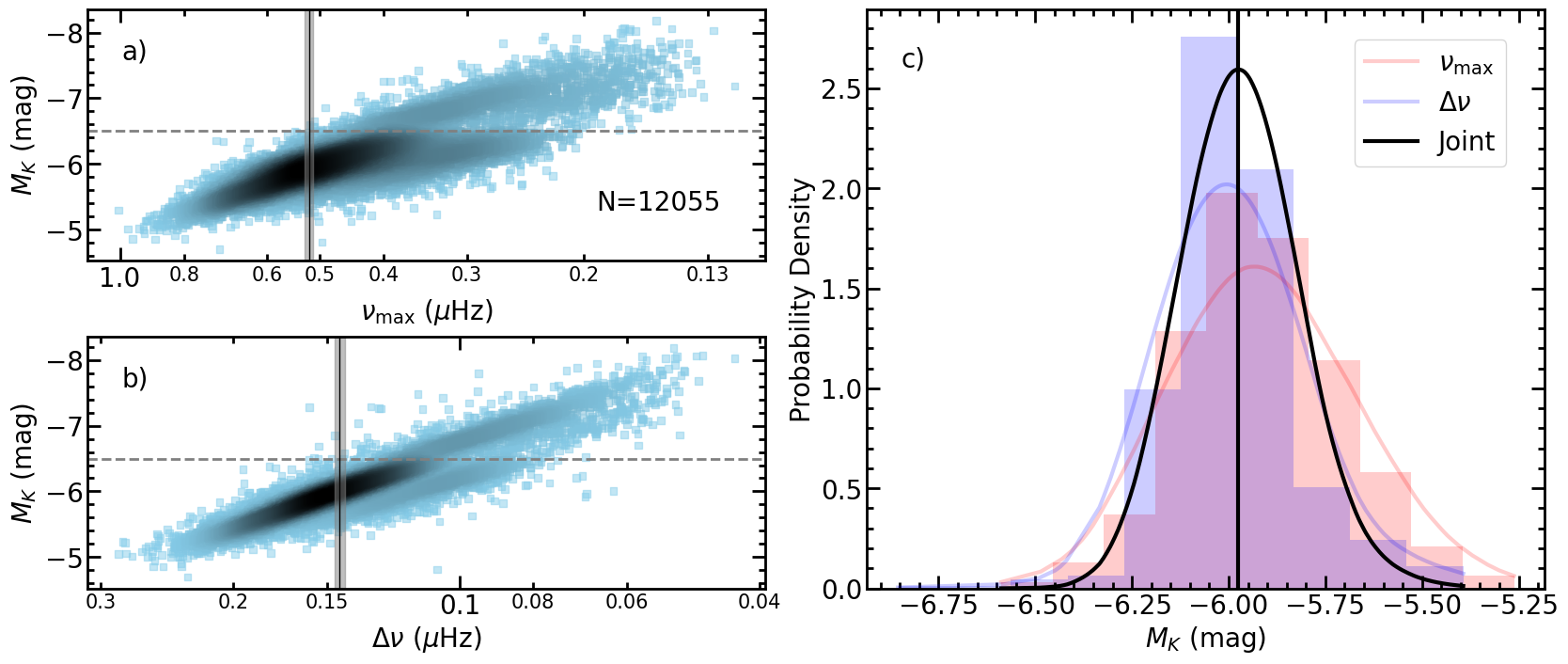}
    \caption{Schematic diagram for calculating the joint PDF of \mk. \textbf{Panel a)} shows the relation between \numax\ and \mk\ for our full sample of stars in the LMC, while \textbf{Panel b)} shows the relation between \Dnu\ and \mk. The black lines and the surrounding gray-shaded regions indicate the \numax\ and \Dnu\ measurements, along with their uncertainties, for OGLE-BLG-LPV-032896 in the BLG. The gray horizontal dashed line marks the \mk\ value corresponding to the tip of the RGB. \textbf{Panel c)} shows the distribution of \mk\ in the LMC derived from this star’s \numax, \Dnu, and their uncertainties, along with the corresponding PDFs calculated using KDE: red for \numax\ and blue for \Dnu. The black line represents the joint PDF obtained by multiplying and normalizing the individual PDFs. The vertical black line represents the \mk\ corresponding to the peak of the PDF. This joint PDF is subsequently used to calculate \mk\ and its associated uncertainties (see Section~\ref{sec:calc_mk}).}
    \label{fig:joint_mk}
\end{figure*}

\begin{figure}[htbp]
    \centering
    \includegraphics[width=1.\linewidth]{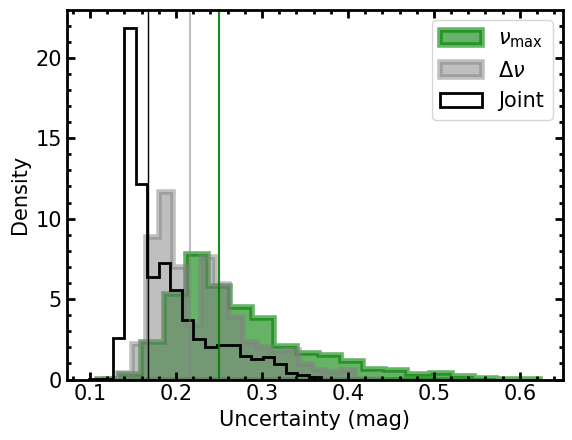}
    \caption{Uncertainty distributions of \mk\ determined using \numax\ (blue), \Dnu\ (green), and the combination of \numax\ and \Dnu\ (black) for all the stars in our sample in the BLG. The black, yellow, and blue vertical lines indicate the median uncertainties of the entire star sample.}
    \label{fig:MK_err}
\end{figure}

\begin{figure}[!htbp]
    \centering
    \includegraphics[width=1. \linewidth]{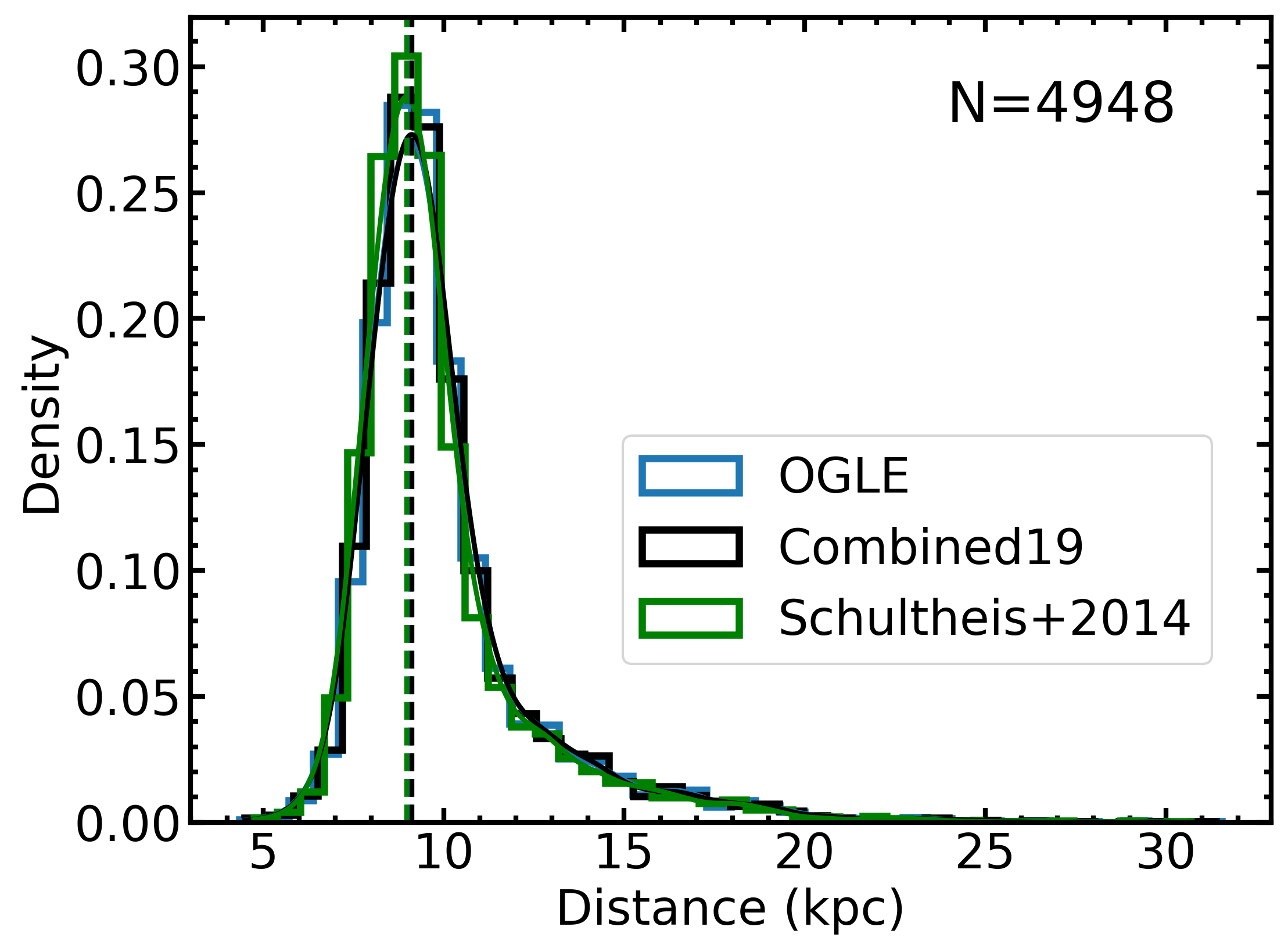}
    \caption{The distance distributions for 4,948 BLG stars based on the OGLE extinction map (blue, \citealt{2012A&A...543A..13G, 2013ApJ...769...88N}), Combined19 extinction map (black, \citealt{2016ApJ...818..130B}), and Schultheis extinction map (green, \citealt{2014A&A...566A.120S}). Solid lines represent the probability density functions (PDFs) obtained through Gaussian kernel density estimation (KDE), while dashed lines indicate the distance modulus corresponding to the peaks of the PDFs.}
    \label{fig:D}
\end{figure}

\subsection{Deriving Absolute Magnitudes \mk\ for BLG stars} \label{sec:calc_mk}

Based on the results from the LMC sample, we derived the \numax-\mk\ and \Dnu-\mk\ relations. Using these relations, we can estimate the absolute magnitude \mk\ for each star in the BLG sample. By combining this with the apparent magnitude and extinction, we can then infer the distance to each star in the BLG sample.

The method used to calculate \mk\ is similar to that described in \citet{tabur2010period} and \citet{hey2023far}. Figure~\ref{fig:joint_mk} illustrates the process of calculating \mk\ for a BLG star. We first derived the \mk\ distribution from a subset of LMC stars whose \numax\ values match that of the BLG star (vertical black line in panel a) within its uncertainty (represented by the grey-shaded region). This \mk\ distribution is shown in red in panel c). Similarly, we obtained another \mk\ distribution by substituting \numax\ with \Dnu, depicted in blue also in panel c). Next, we applied the kernel density estimation (KDE) method to compute the probability density functions (PDFs) of \mk\ for each seismic parameter. These PDFs are represented by the light red and light blue lines in panel c). Silverman’s rule of thumb is applied to automatically select the bandwidth, effectively balancing the degree of smoothing to avoid both over-smoothing and under-smoothing, thus capturing the distribution features more accurately. Finally, we multiplied these PDFs and normalized them to obtain the joint PDF, shown as the solid black line in panel c) of Figure~\ref{fig:joint_mk}. The \mk\ corresponding to the peak of the joint PDF represents the best estimate for this star ($M_{K,\ \text{best}}$), and the uncertainty in the estimate of \mk\ can also be derived from this PDF, defined as follows:

\begin{equation}
    \sigma_{M_{K,\ \text{best}}} = \sqrt{\frac{\sum_i p_{\text{joint},i}(M_{K,i}-M_{K,\ \text{best}})^2}{\sum_i p_{\text{joint},i}}},
\end{equation}

\noindent where $p_{\text{joint}, i}$ denotes the probability density value from the joint PDF, and $M_{\text{K}, i}$ represents the \mk\ value corresponding to each $p_{\text{joint}, i}$.

Since the relation between the two seismic parameters and \mk\ can exhibit two branches in certain regions ($\sim$ 0.240 to 0.340 $\mu$Hz for \numax\ and 0.085 to 0.110 $\mu$Hz for \Dnu),

Since the relation between the two seismic parameters and \mk\ can exhibit two branches in certain regions ($\sim$ 0.240 to 0.340 $\mu$Hz for \numax\ and 0.085 to 0.110 $\mu$Hz for \Dnu), some stars may exhibit a bimodal distribution in the joint PDF of \mk. This phenomenon is likely due to the fact that AGB and RGB stars do not fully overlap in the period-luminosity relation. Specifically, in sequence A, there is a noticeable offset between the regions occupied by RGB and AGB stars. To verify this bimodality, we performed a peak-finding analysis to identify the presence of two distinct peaks in the distribution. As a result, we excluded these stars from the analysis. Additionally, if the \mk\ values for a given parameter contain fewer than thirty data points, the PDF estimation is considered unreliable, and such stars were excluded as well. In summary, we identified 5,370 stars with an unimodal joint PDF, 933 with a bimodal joint PDF, and 88 for which the joint PDF estimate was deemed unreliable. It is worth noting that using both \numax\ and \Dnu\ can significantly improve precision compared to using either indicator alone. As demonstrated in Figure~\ref{fig:MK_err}, the median precision of \mk\ is 0.25 mag for the case of using \numax\ as the indicator. The median precision is improved to 0.22 mag for \Dnu, and 0.17 mag while both two indicators are at play.

To estimate the extinction ($A_{K_\text{s}}$) for BLG stars, we considered three extinction maps: the map derived from OGLE\footnote{\href{https://ogle.astrouw.edu.pl/}{https://ogle.astrouw.edu.pl/}} (\citealt{2012A&A...543A..13G, 2013ApJ...769...88N}), \texttt{Combined19} from \mbox{\texttt{mwdust}} \citep{2016ApJ...818..130B}, and the map by \citet{2014A&A...566A.120S}. The extinction coefficient ($R_{K_s}$) is adopted as 0.689, according to \citet{1989ApJ...345..245C}. Similar to the description in Section~\ref{sec:target}, we cross-matched with the 2MASS database to obtain high-quality photometric measurements of $K_\text{s}$-band apparent magnitudes ($K_{\text{mag}}$). For the OGLE map, we also selected extinction measurements with a quality factor of \texttt{QF=0}.

We obtained distance moduli ($\mu$) for 4,948 BLG stars consistently across all three extinction maps. These were then converted to distances in kiloparsecs (kpc) and are shown in Figure~\ref{fig:D}. To estimate the uncertainty in the distance modulus ($\sigma_{\mu}$) for each star, a Monte Carlo approach is adopted. The best estimates of apparent magnitude, absolute magnitude, and extinction define the mean vector, while their respective uncertainties serve as standard deviations. Since none of the three extinction maps provide uncertainty estimates for extinction values, the standard deviation of the extinction values from the three maps is used as the uncertainty estimate for individual stars. These uncertainties are then incorporated into a covariance matrix, which is used to define a multivariate normal distribution. We draw 10,000 random samples from this distribution for each star and compute the median absolute deviation of the resulting distance moduli to quantify $\sigma_{\mu}$. The peak of the resulting distance distribution corresponds to estimated distances to the Galactic center of $9.07 \pm 0.33_\text{stat} \pm 0.11_\text{sys}$, $9.11 \pm 0.25_\text{stat} \pm 0.11_\text{sys}$, and $9.08 \pm 0.27_\text{stat} \pm 0.11_\text{sys}$ kpc, based on the extinction maps from OGLE (\citealt{2012A&A...543A..13G,2013ApJ...769...88N}), \texttt{Combined19} \citep{2016ApJ...818..130B}, and \citet{2014A&A...566A.120S}, respectively. The best-estimated value and its associated statistical uncertainty are derived using the bootstrapping technique. Specifically, we performed 10,000 iterations, in each of which the original dataset was resampled with replacement. For each resampled dataset, the distance corresponding to the peak of the resulting distance distribution was calculated. The median of the distances from all iterations was adopted as the best-estimated value, and the standard deviation was taken as the statistical uncertainty in the Galactic center distance. The systematic uncertainty arises from the uncertainty in the distance modulus of the LMC. These small variations indicate that the choice of extinction map has a negligible impact on the derived distances. However, the distances to the Galactic center derived from the three extinction maps all exhibit a significant deviation from the value obtained by modeling the orbit of the star S2 around Sgr A$^*$, which yields a distance of $8275 \pm 9_{\text{stat}} \pm 33_{\text{sys}}$ pc (\citealt{2021A&A...647A..59G}). We have thoroughly checked our results, including the measurements of \numax\ and \Dnu, and have also tested various methods for calculating \mk, such as using machine learning and interpolation techniques. The results we obtained were consistent with our current findings. We were unsure of the reasons behind this discrepancy, it may result from the significant environmental differences between the BLG and LMC, causing the \numax-\mk\ and \Dnu-\mk\ relations from the LMC to be unsuitable for direct use with the BLG sample, which warrants further investigation in the future. The \mk\ values, along with the mean distance moduli derived from the three extinction maps for the BLG stars, were summarized in Table~\ref{tab:starlist_blg}.

\section{Conclusions} \label{sec:conclusion}

In this work, we present an asteroseismic analysis of 53,273 OSARGs in the LMC and 23,266 in the BLG, using data from OGLE-II, OGLE-III \citep{soszynski2009optical, soszynski2013optical}, and OGLE-IV \citep{udalski_ogle-iv_2015}. By modeling the OGLE light curves in the time domain with a Gaussian Process method, we mitigated the influence of observation gaps, which can significantly affect analyses in the frequency domain when gaps are non-negligible.

Our models consist of two components: one characterizing noise, granulation, and oscillations, which provides \numax, and the other modeling three pulsation modes, as three orders of pulsation modes are evident in many OSARGs. This second component allows us to measure the average frequency separation, denoted as \Dnu. From our analysis, we derived \numax\ and \Dnu\ values for 12,055 stars in the LMC and 6,391 stars in the BLG. The \numax\ values range from 0.12 to 1.01 $\mu$Hz.

The relation between \numax\ and \Dnu\ measured in this work closely matches that of \kepler\ red giants \citep{yu2018asteroseismology, yu2020asteroseismology}, suggesting that OSARGs exhibit solar-like oscillations, consistent with earlier studies \citep[e.g.,][]{2010A&A...524A..88D, takayama_pulsation_2013, yu2020asteroseismology, trabucchi2024selfexcitedpulsationsinstabilitystrip}. Through our analysis of LMC stars, we demonstrate that both \numax\ and \Dnu\ serve as robust luminosity indicators in this work. Compared to the primary period (\p1), these parameters offer significant advantages as they do not exhibit the prominent double sequences characteristic of \p1. The scatter in the \numax-\mk\ relation is 0.27 mag, while that in the \Dnu-\mk\ relation is 0.21 mag.

Leveraging the relations between seismic parameters and \mk\ derived from LMC stars, we estimated \mk\ for BLG stars. By incorporating extinction (\citealt{2012A&A...543A..13G, 2013ApJ...769...88N, 2016ApJ...818..130B,2014A&A...566A.120S}), we determined distances for 4,948 BLG stars. The peak of the resulting distance distribution corresponds to a distance to the Galactic center of approximately 9.1 kpc, which is larger than the values reported in previous studies (e.g., $\sim $ 8.3 kpc; \citealt{{2021A&A...647A..59G}}). This discrepancy suggests that the \numax-\mk\ and \Dnu-\mk\ relations established for LMC stars may not be directly applicable to BLG stars.

\section*{Acknowledgements}


We sincerely thank the OGLE team for generously providing the OGLE-IV data, which offers exceptional observational duration and photometric precision, making this work possible.\\


\software{astropy \citep{2013A&A...558A..33A},
          CELERITE2 \citep{celerite1, celerite2},
          matplotlib \citep{Hunter:2007},
          numpy \citep{harris2020array},
          scikit-learn \citep{scikit-learn},
          scipy \citep{2020SciPy-NMeth}
}

\FloatBarrier

\bibliography{refs}
\bibliographystyle{aasjournal}

\appendix 
\restartappendixnumbering

\section{Examples of PSD obtained through GP modeling.} \label{appendixA}

Here, to further demonstrate the effectiveness of Gaussian Processes (GP) in the asteroseismic analysis of ground-based telescope data, we presented the frequency-domain representations of the fitting results for three randomly selected stars from both the LMC (Figure~\ref{fig:lmc_psd}) and the BLG (Figure~\ref{fig:blg_psd}), using the \mnumax\ and \mdnu\ models, respectively.

\renewcommand{\tablename}{Figure}
\begin{table*}[h!]
    \centering
    \begin{tabular}{cc} %

        \includegraphics[width=.47\textwidth]{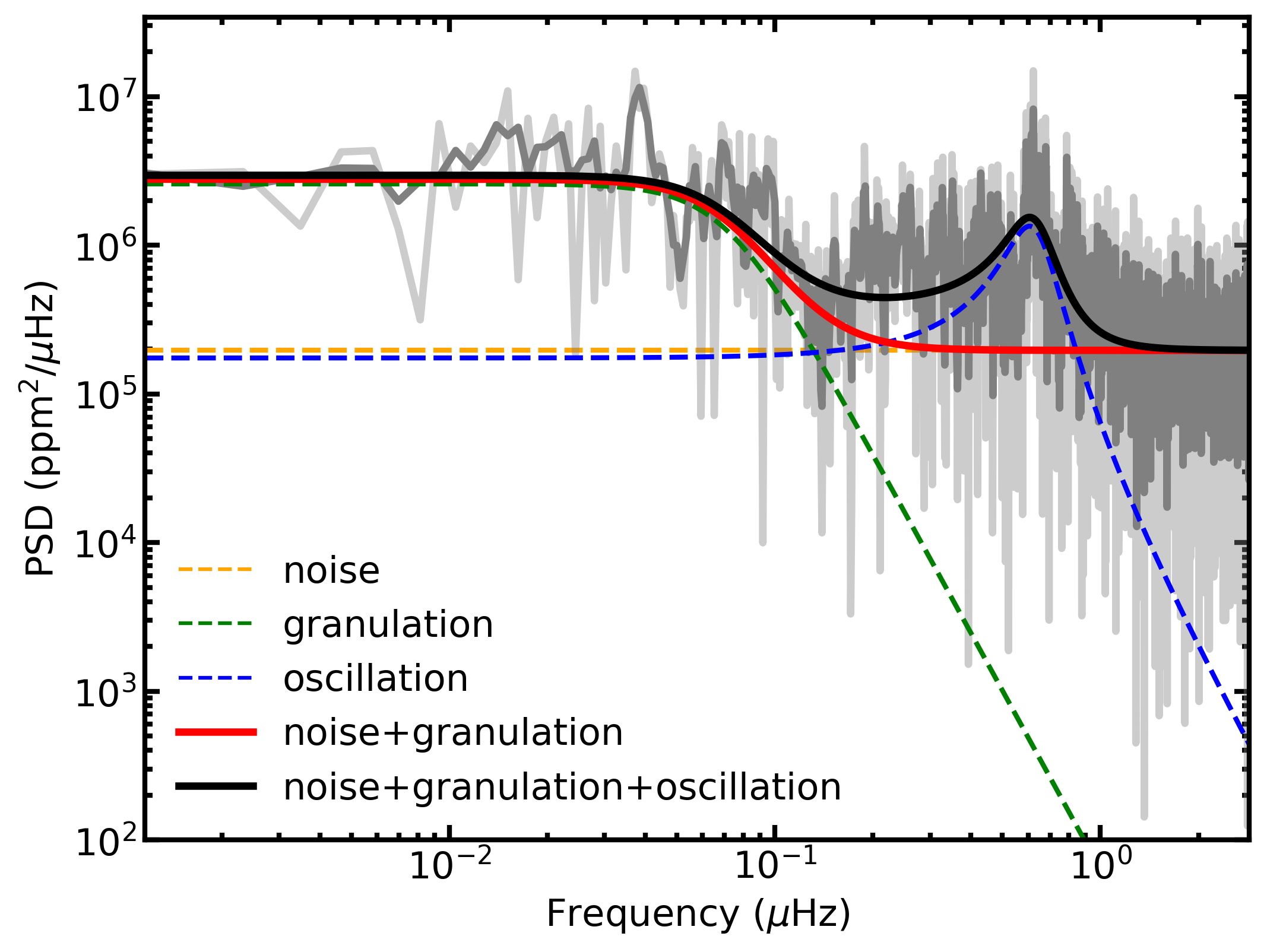} & 
        \includegraphics[width=.47\textwidth]{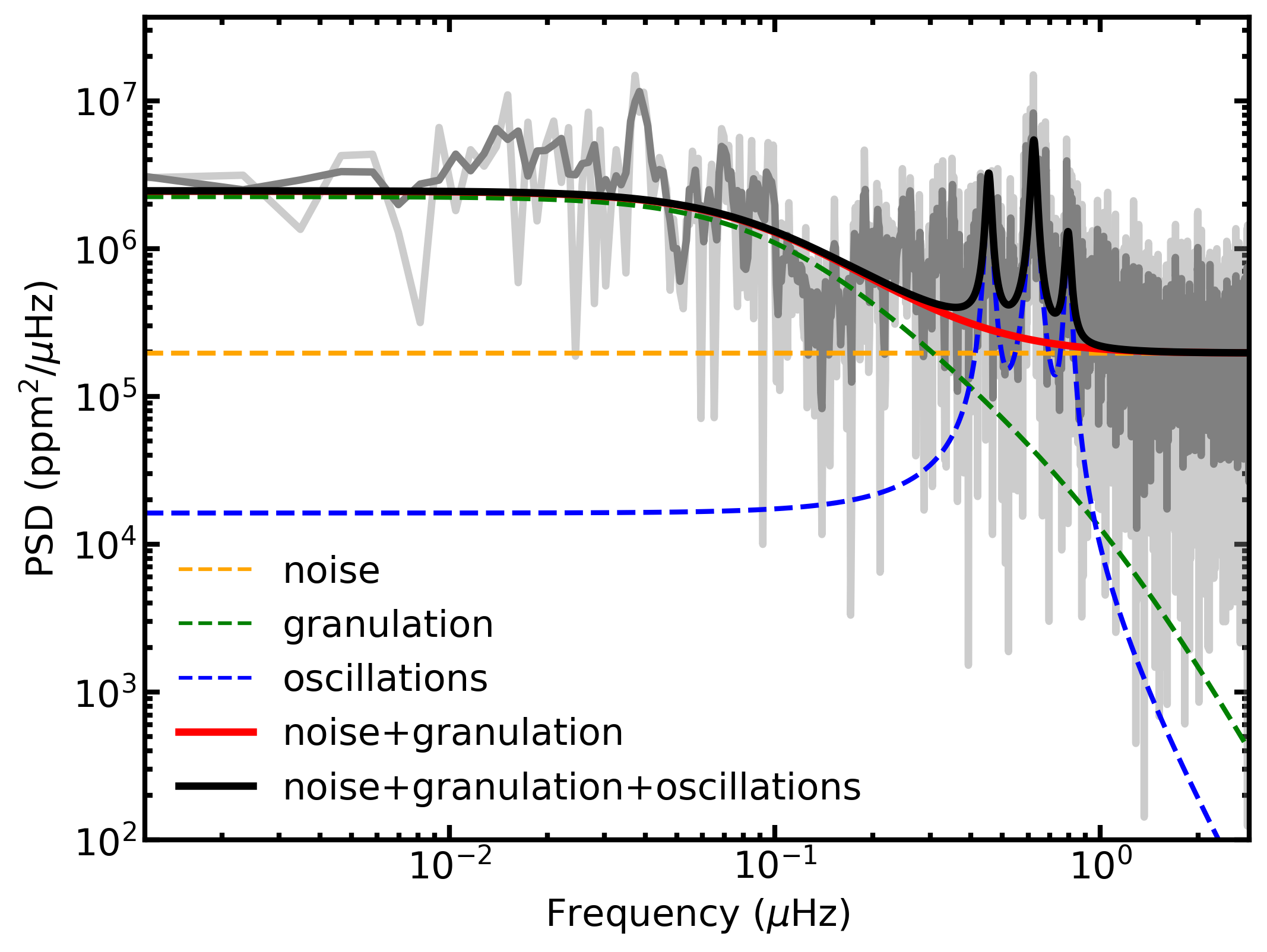} \\

        \includegraphics[width=.47\textwidth]{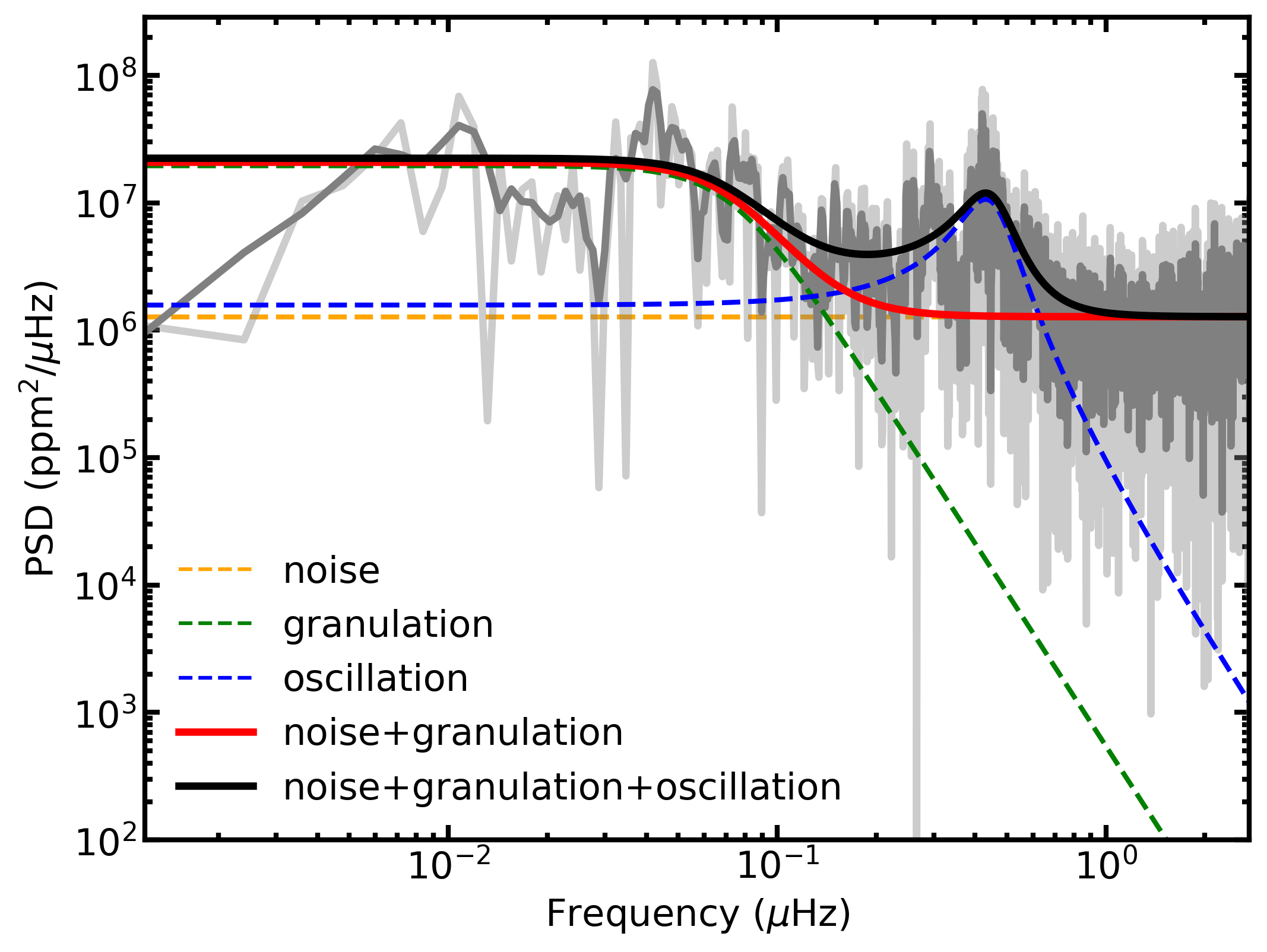} & 
        \includegraphics[width=.47\textwidth]{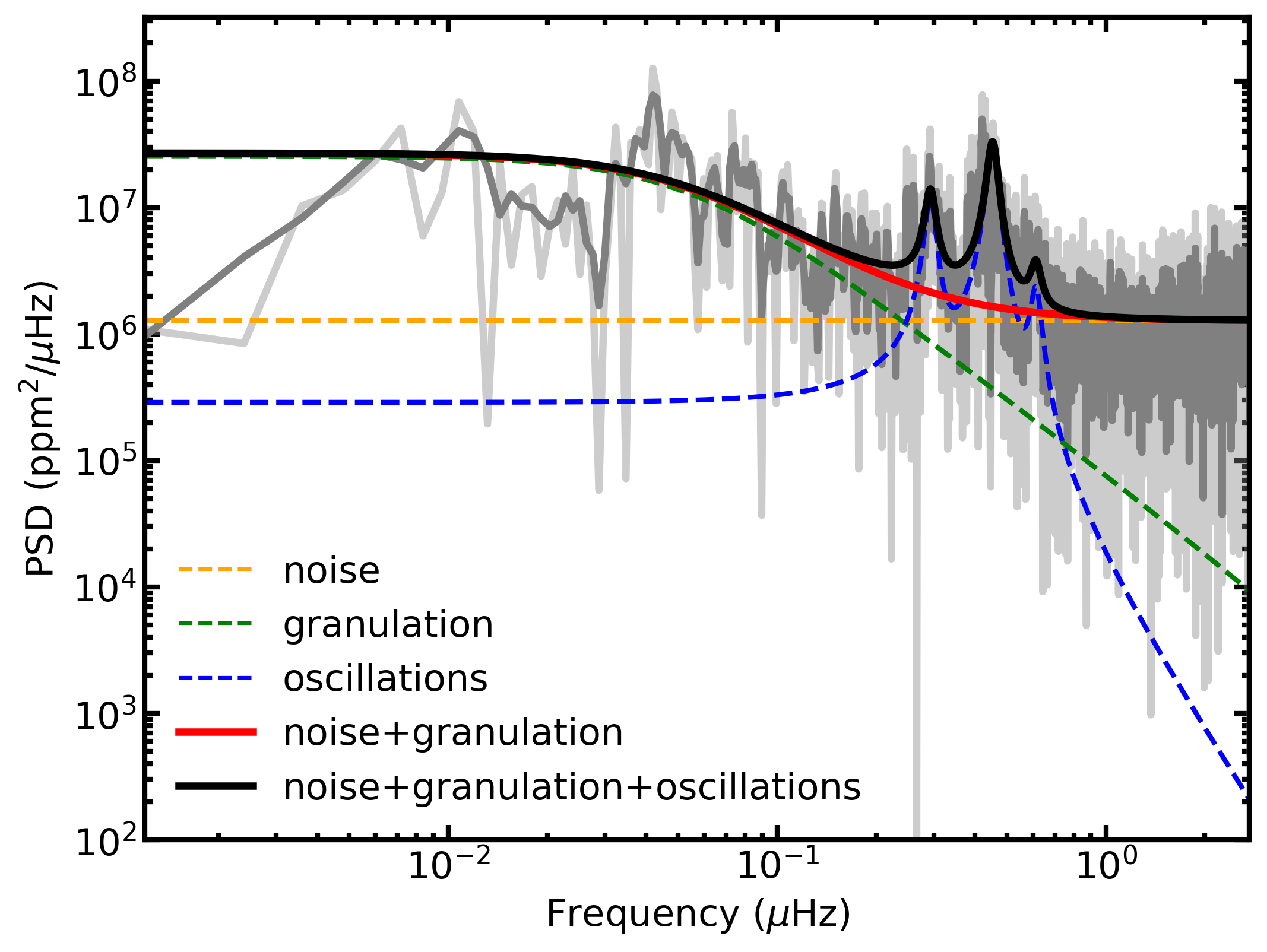} \\

        \includegraphics[width=.47\textwidth]{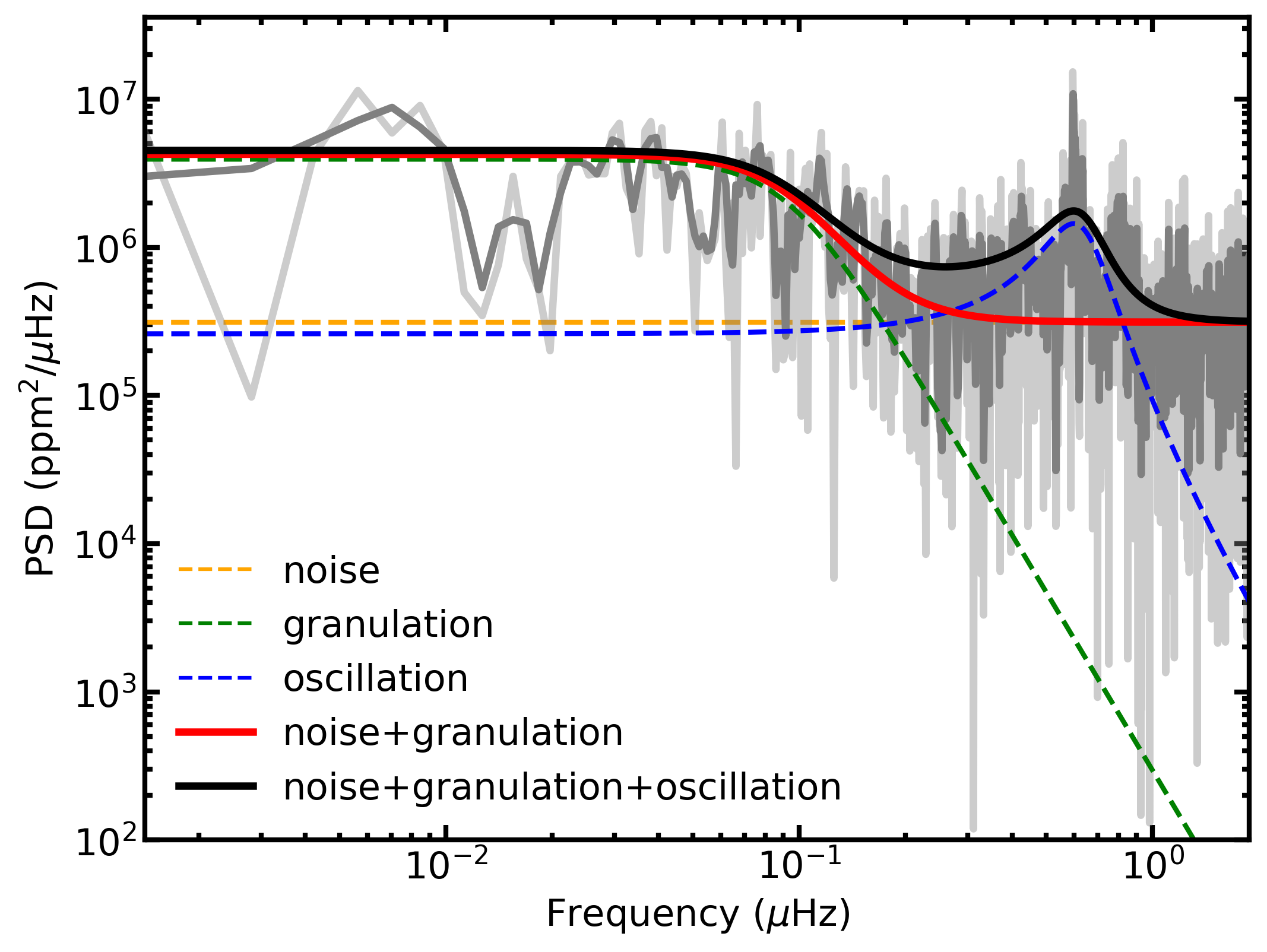} & 
        \includegraphics[width=.47\textwidth]{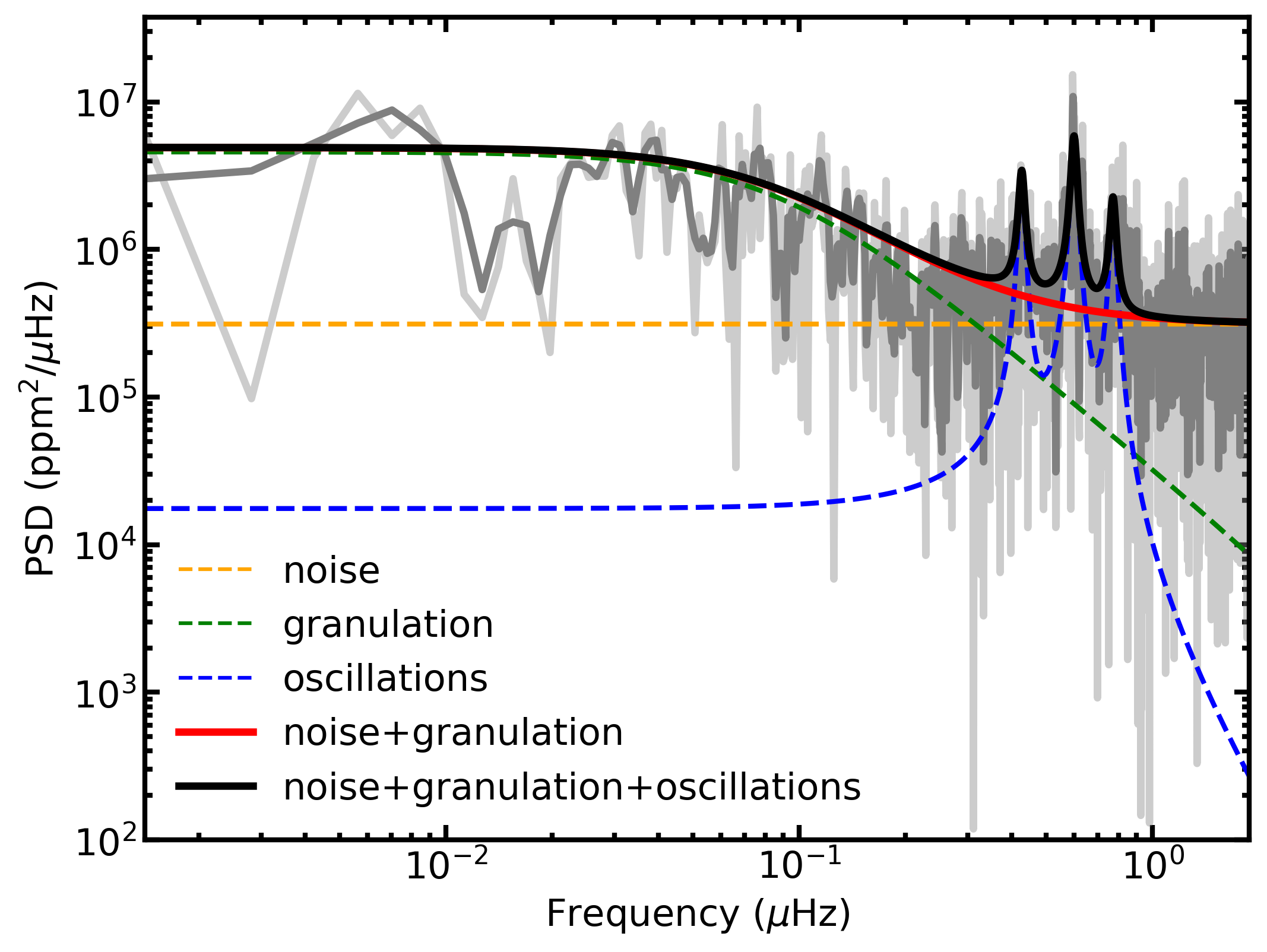} \\
        
    \end{tabular}
    \caption{Frequency-domain representations of the light curve fittings for three LMC stars through GP, using the \mnumax\ model (left panel) and the \mdnu\ model (right panel). From top to bottom, the stars are \texttt{OGLE-LMC-LPV-68008}, \texttt{OGLE-LMC-LPV-81553}, and \texttt{OGLE-LMC-LPV-90842}.}
    \label{fig:lmc_psd}
\end{table*}

\begin{table*}[h!]
    \centering
    \begin{tabular}{cc}

        \includegraphics[width=.47\textwidth]{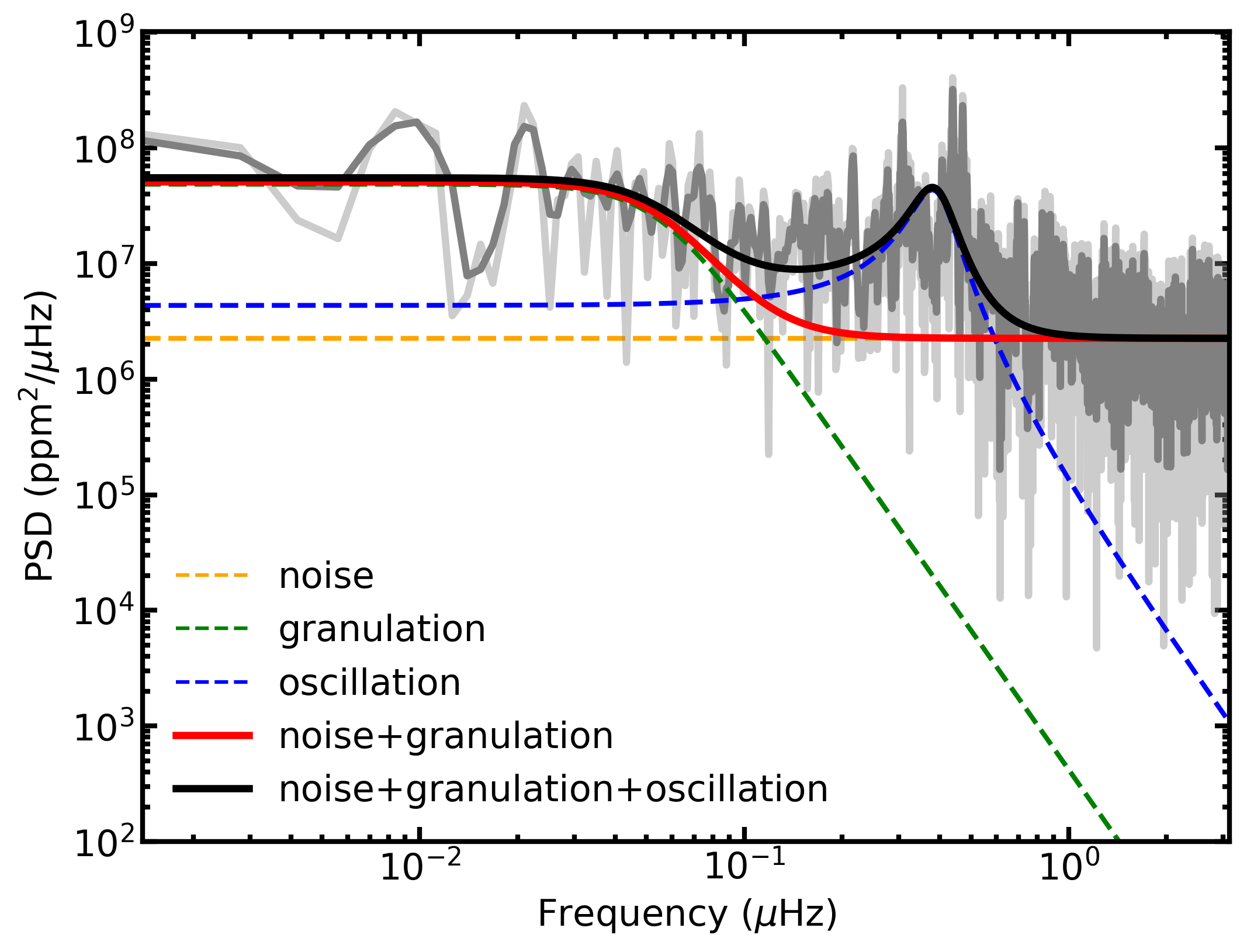} & 
        \includegraphics[width=.47\textwidth]{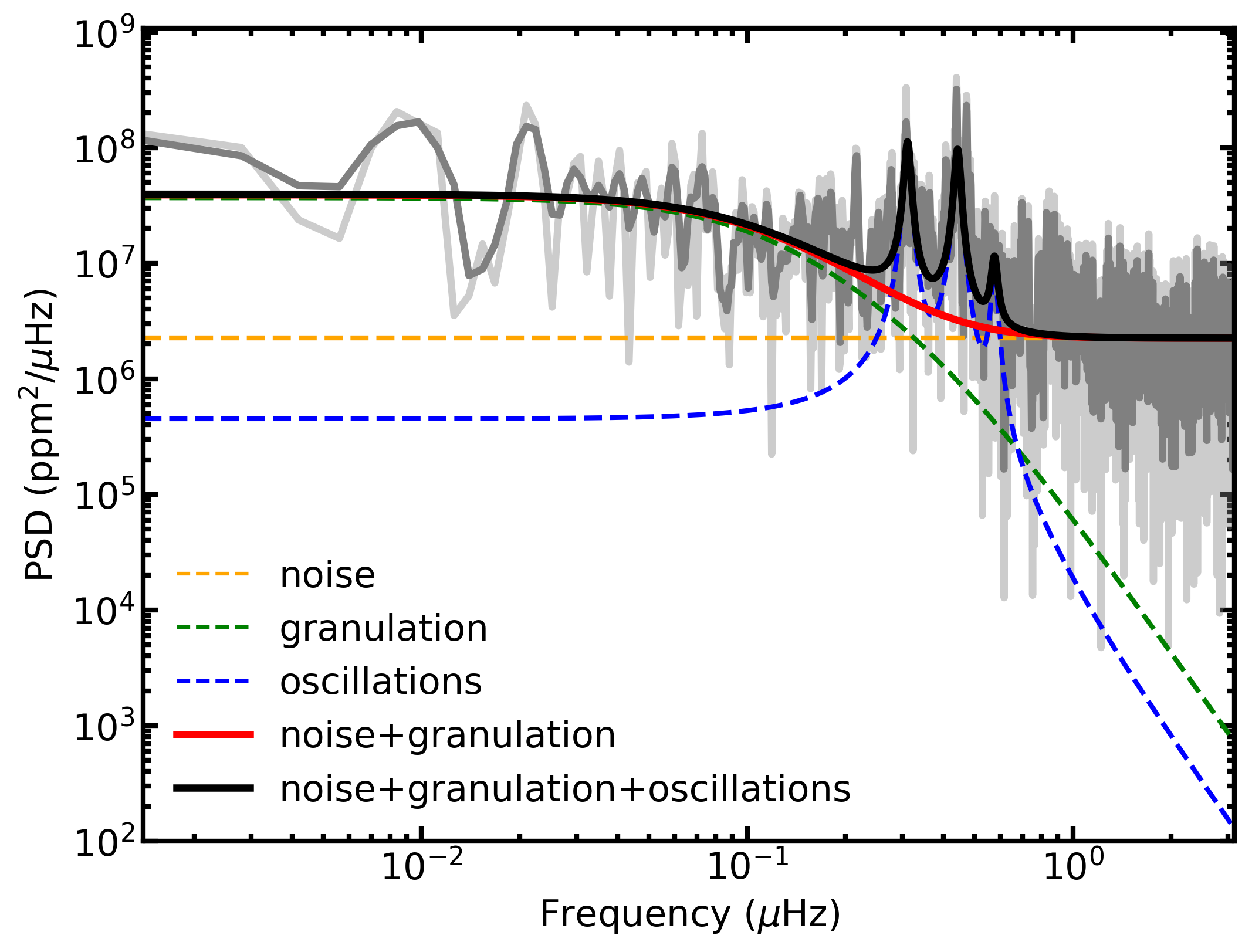} \\

        \includegraphics[width=.47\textwidth]{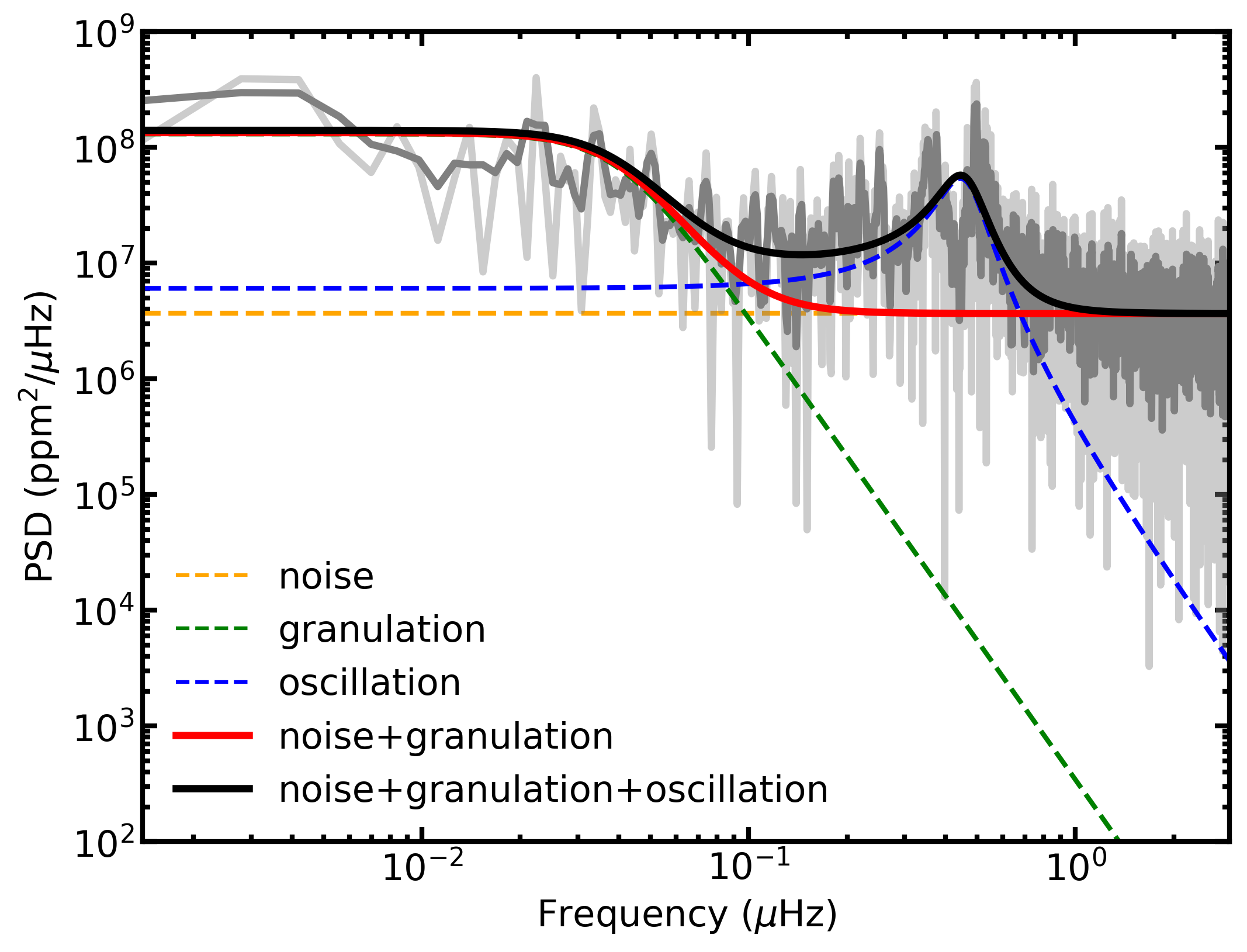} & 
        \includegraphics[width=.47\textwidth]{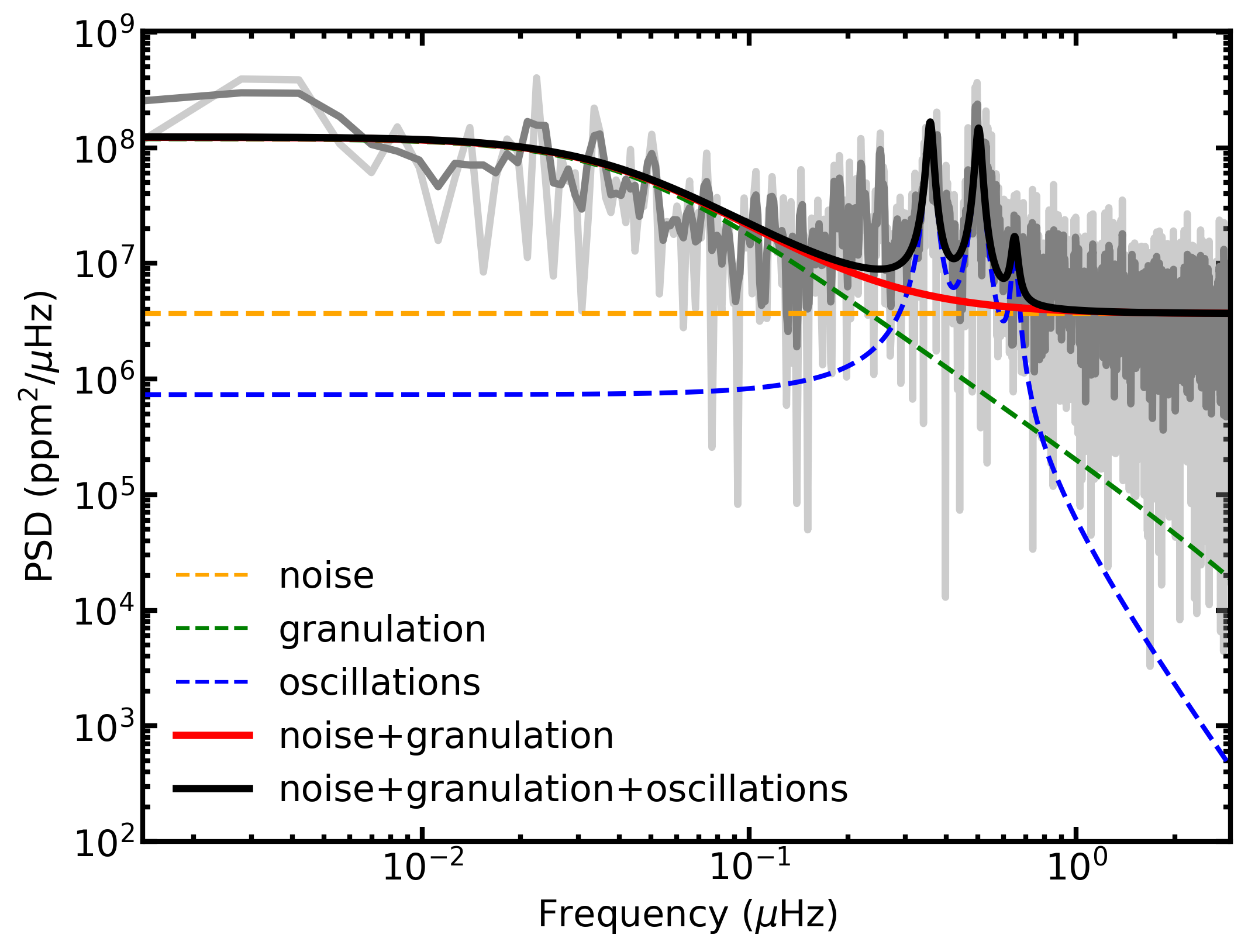} \\

        \includegraphics[width=.47\textwidth]{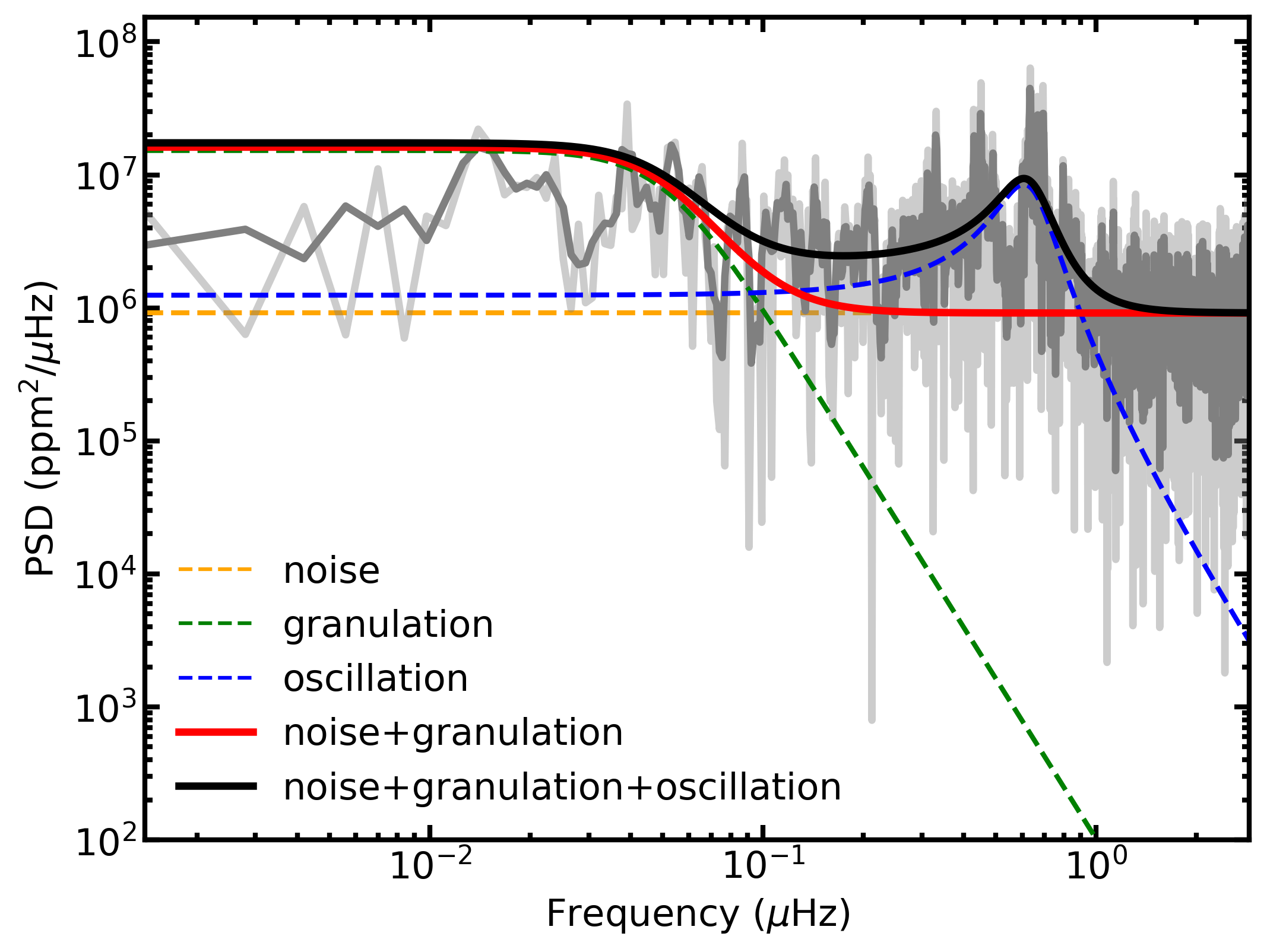} & 
        \includegraphics[width=.47\textwidth]{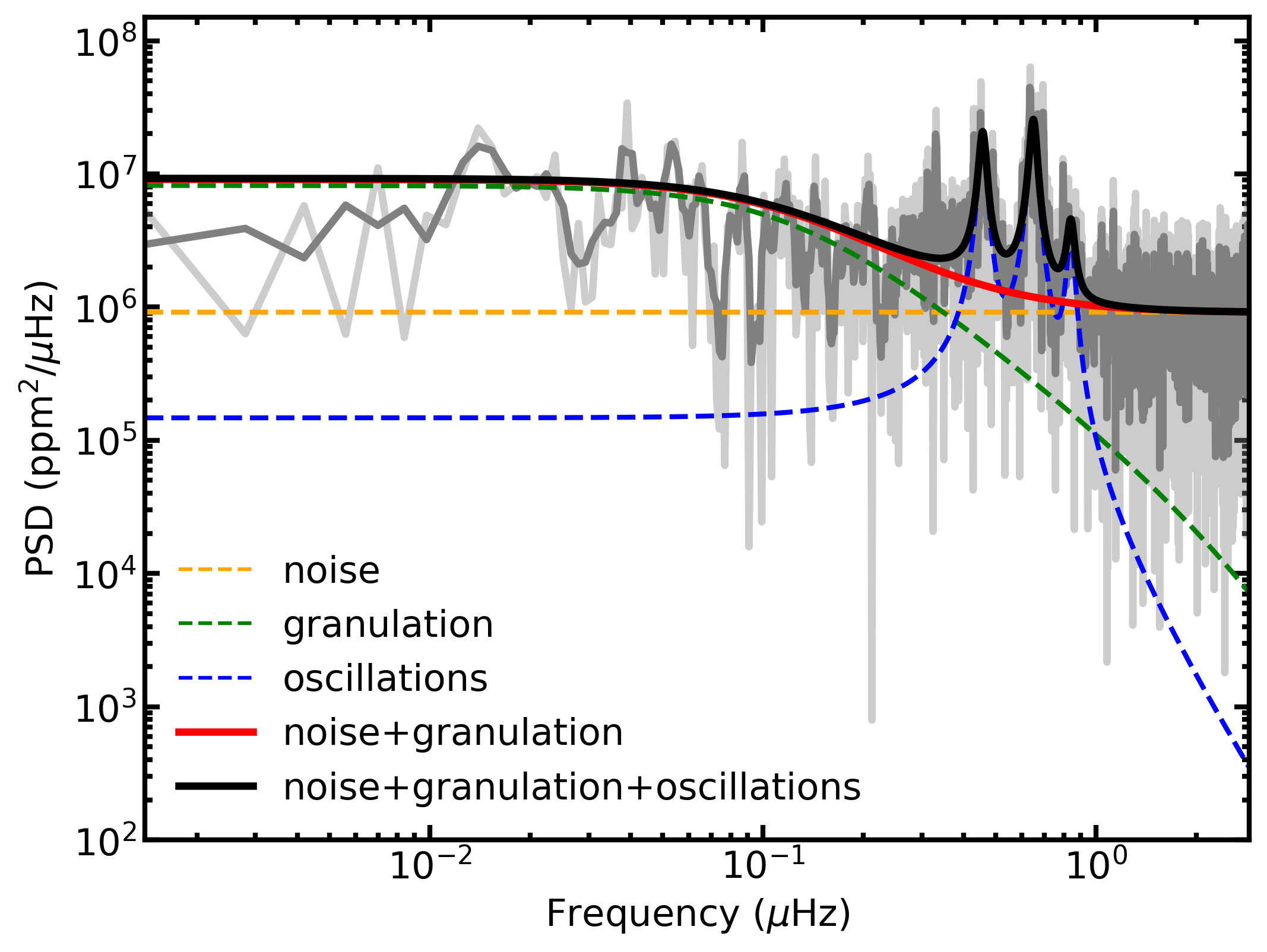} \\
        
    \end{tabular}
    \caption{Same as Figure~\ref{fig:lmc_psd}, but for BLG stars. From top to bottom, the stars are \texttt{OGLE-BLG-LPV-006667}, \texttt{OGLE-BLG-LPV-067574}, and \texttt{OGLE-BLG-LPV-192962}.}
    \label{fig:blg_psd}
\end{table*}

\end{document}